\documentclass[superscriptaddress,preprint,eqsecnum,amsmath,aps,prb]{revtex4}
\usepackage{graphicx}

\begin{document}

\title{Linear optical properties of solids within the full-potential 
linearized augmented planewave method}

\author{C. Ambrosch-Draxl}
\affiliation{Institut f\"{u}r Theoretische Physik, Universit\"{a}t
  Graz, Universit\"{a}tsplatz 5, A-8010 Austria}
\author{J. O. Sofo}
\affiliation{Institut f\"{u}r Theoretische Physik, Universit\"{a}t
  Graz, Universit\"{a}tsplatz 5, A-8010 Austria}
\affiliation{Department of Physics and Materials Research Institute,
  The Pennsylvania State University, 104 Davey Lab. PMB\#172,
  University Park PA 16802, USA} 
\date{\today}

\begin{abstract}
We present a scheme for the calculation of linear optical properties
by the all-electron full-potential linearized augmented planewave
(LAPW) method. A summary of the theoretical background for the
derivation of the dielectric tensor within the random-phase
approximation is provided together with symmetry considerations and
the relation between the optical constants. The momentum matrix
elements are evaluated in detail for the LAPW basis, and the
interband as well as the intraband contributions to the dielectric
tensor are given. Results are presented for the metals aluminum and
gold, where we crosscheck our results by sumrules. We find that the
optical spectra can be extremely sensitive to the Brillouin zone
sampling. For gold, the influence of relativistic effects on the
dielectic function is investigated. It is shown that the
scalar-relativistic effect is much more important than spin-orbit
coupling. The interpretability of the Kohn-Sham eigenstates in terms
of excited states is discussed.
\end{abstract}

\maketitle

\narrowtext

\section{Introduction}

Optical properties of solids are a major topic, both, in basic
research as well as for industrial applications. While for the former
origin and nature of different excitation processes is of fundamental
interest, the latter can make use of them in many opto-electronic
devices. These wide interests require experiment and theory to go hand
in hand, and thus asks for reliable theoretical concepts.

In solid state theory, the most successful method for the calculation
of materials properties is density functional theory (DFT). Lattice
properties like equilibrium volumes and lattice parameters, atomic
positions, phonon frequencies and elastic constants differ from their
experimental counterparts by a few percent only. While, however, such
ground state (GS) properties, which are based on the calculation of
the total energy, are described very reliably, the treatment of
excited states is not rigorously justified. The main reasons are that
the Hohenberg-Kohn \cite{Kohn:64} theorem is exact only for the GS,
and that the Kohn-Sham eigenstates \cite{Sham:65} should not be
interpreted as single-electron states. Moreover, approximations have
to be made also in the ground state calculations for describing
exchange and correlation effects. As a consequence, the band gap
problem for semiconductors is one of the most intriguing problems in
the field.

Nevertheless the interpretation of the KS states in terms of excited
states has been successful for a variety of materials. Moreover, it
has been claimed that the KS wavefunctions hardly differ from the
many-body wavefunctions.\cite{Hybertsen,Levine} It has to be pointed
out that the calculation of optical properties does not go beyond the
interpretation of KS eigenvalues in terms of the band structure.  In
this context, it may be hard to distinguish whether the shortcomings
are due to the over-interpretation of GS properties or to the
approximations used to solve the KS equations, e.g. the local density
approximation (LDA). To gain more insight, a series of results is
needed where no further approximations are introduced when solving the
KS equations.

To this extent we have chosen the full potential linearized augmented
planewave (LAPW) method where no shape approximation for the potential
is made.  We have developed the formalism of treating optical
properties within the random phase approximation (RPA) taking into
account inter- as well as intraband contributions. In terms of
applications we focus on metallic cases for the following reasons. (i)
Metals don't suffer from the band gap problem; (ii) metals are quite
well described by the LDA, and (iii) the RPA is justified due to the
effective screening.

The paper is organized as follows: In Section \ref{theory}, the
optical response within the RPA is discussed, where a short comparison
to time-dependent density functional theory (TDDFT) is made. Symmetry
considerations, and the description of other optical constants
obtained via the Kramers--Kronig relations follow hereafter. Section
\ref{optic_LAPW} is dedicated to the treatment of the optical response
within the LAPW method. First, the basic definitions of the needed
quantities, are given, then the momentum matrix elements are derived
within the LAPW basis. The next section contains results obtained for
the elemental metals aluminum and gold. 
Finally, conclusions are drawn on our results for the two metallic cases.
The extension of the method with localized basis functions 
(local orbitals) is described in the Appendix together with some
mathematical relations which are useful in the derivation of the formulas.

\section{Theoretical background}
\label{theory}

\subsection{Optical response}
\label{sub_response}

The optical properties of solids are given by the response of the
electron system to a time-dependent electromagnetic perturbation
caused by the incoming light. As such, the calculation of these
properties is reduced to the calculation of a response function that
is the complex dielectric tensor or equivalently the polarizability.
An exact expression for it is of course not known and one has to
resort to the usual techniques of many-body perturbation theory to
derive approximations.  The first of these is the RPA.

A particular case of the general expression for the dielectric
function in the RPA is the well known Lindhard formula,
\cite{Ashcroft_p344}
\begin{equation}
  \label{eq:lindhard}
  \epsilon(\mathbf{q},\omega) = 1+ \frac{2 v(\mathbf{q})}{\Omega_c}
  \sum_{\mathbf{k}} \frac{f_0(\varepsilon_{\mathbf{k}+\mathbf{q}}) -
  f_0(\varepsilon_{\mathbf{k}})}{\varepsilon_{\mathbf{k}+\mathbf{q}} - 
  \varepsilon_{\mathbf{k}}-\omega}\;,
\end{equation}
where 
\begin{equation}
  \label{eq:coulq}
  v(\mathbf{q})=\frac{4\pi e^2}{\left|\mathbf{q}\right|^2}
\end{equation}
is the Coulomb
interaction, $\Omega_c$ the unit cell volume, $f_0$ the Fermi
distribution function, and $\varepsilon_\mathbf{k}$ the single
particle energy. The factor 2 comes from the summation over the spin.
This corresponds to a free electron gas in the Hartree approximation,
namely, the electron-electron interaction is reduced to the
interaction of each electron with a homogeneous self-consistent
field. Many equivalent ways of deriving this expression have been
given in the literature.\cite{Adler,Wiser} 

According to Hedin \cite{Hedin65} the dielectric function is defined
as
\begin{equation}
  \label{eq:hedin}
\epsilon(\mathbf{r},t;\mathbf{r}^\prime,t^\prime)=\delta(\mathbf{r}-\mathbf{r}^\prime)\delta(t-t^\prime)-\int
  P(\mathbf{r},t;\mathbf{r}^{\prime\prime},t^{\prime})
  v(\mathbf{r}^{\prime\prime}-\mathbf{r}^\prime) d\mathbf{r}^{\prime\prime} 
\end{equation}
where $v$ is the bare Coulomb interaction and $P$ is the polarization
propagator. The simplest approximation for $P$ is the RPA that
corresponds to the Hartree approximation for the one particle Green
function $G^0(\mathbf{r},t;\mathbf{r}^\prime,t^\prime)$. In this
approximation, the polarization propagator $P^0$ is given by
\begin{equation}
  \label{eq:p0}
  P^0(\mathbf{r},t;\mathbf{r}^\prime,t^\prime)=-i \hbar
  G^0(\mathbf{r},t;\mathbf{r}^\prime,t^\prime) 
  G^0(\mathbf{r}^\prime,t^\prime;\mathbf{r},t)
\end{equation}
For the case of a solid with translational symmetry we can Fourier
transform the equation for the dielectric function
Eq.~(\ref{eq:hedin}) to obtain its expression as a tensor in the
reciprocal lattice vectors:
\begin{equation}
  \label{eq:dieltens}
  \epsilon_{\mathbf{G},\mathbf{G}^\prime}(\mathbf{q},\omega) =
  \delta_{\mathbf{G},\mathbf{G}^\prime} - v(\mathbf{q}+\mathbf{G}) P^0_{\mathbf{G},\mathbf{G}^\prime}(\mathbf{q},\omega)\;,
\end{equation}
where $P^0_{\mathbf{G},\mathbf{G}^\prime}(\mathbf{q},\omega)$ is the
Fourier transform in the lattice of the bare polarizability defined in
Eq.~(\ref{eq:p0}). It is given by
\begin{equation}
  \label{eq:p0q}
  P^0_{\mathbf{G},\mathbf{G}^\prime}(\mathbf{q},\omega) =
  \frac{1}{\Omega_c} \sum_{n',n,\mathbf{k}}
  \frac{f_0(\varepsilon_{n,\mathbf{k}+\mathbf{q}}) - 
 f_0(\varepsilon_{n',\mathbf{k}})}{\varepsilon_{n,\mathbf{k}+\mathbf{q}} 
 - \varepsilon_{n',\mathbf{k}}-\omega} \:
  \left[M_{n',n}^\mathbf{G}(\mathbf{k},\mathbf{q})\right]^\ast \:
  M_{n',n}^{\mathbf{G}^\prime}(\mathbf{k},\mathbf{q}) \; ,
\end{equation}
with the matrix elements $M$ defined as
\begin{equation}
  \label{eq:matelem}
  M_{n',n}^{\mathbf{G}}(\mathbf{k},\mathbf{q}) = \langle n',\mathbf{k}
  \left| e^{-i\,\left(\mathbf{q}+\mathbf{G}\right)\cdot\mathbf{r}}
  \right| n,\mathbf{k}+\mathbf{q} \rangle  \;.
\end{equation}
The expression obtained so far is the dielectric tensor that
connects the total electrostatic potential in the solid, $V$, with the
potential produced only by the external sources, $V^{ext}$, through
the expression
\begin{equation}
  \label{eq:vexttov}
  V^{ext}_{\mathbf{G}}(\mathbf{q},\omega) = \sum_{\mathbf{G}^\prime}
  \epsilon_{\mathbf{G},\mathbf{G}^\prime}(\mathbf{q},\omega)
  V_{\mathbf{G}^\prime}(\mathbf{q},\omega) \;.
\end{equation}
If we denote by $\bar{\epsilon}$ the inverse of the dielectric tensor,
we can invert the relation above to obtain an expression for the total
potential
\begin{equation}
  \label{eq:vtovext}
    V_{\mathbf{G}}(\mathbf{q},\omega) = \sum_{\mathbf{G}^\prime}
    \bar{\epsilon}_{\mathbf{G},\mathbf{G}^\prime}(\mathbf{q},\omega)
    V^{ext}_{\mathbf{G}^\prime}(\mathbf{q},\omega) \;.
\end{equation}
When considering perturbations produced by light, only long wavelength
components are present in the external perturbation potential.
Therefore, it is enough to assume that only the $\mathbf{G}=0$
component is different form zero. In this case the response is
\begin{equation}
  \label{eq:vtovextq0}
    V_{\mathbf{G}}(\mathbf{q},\omega) = 
    \bar{\epsilon}_{\mathbf{G},0}(\mathbf{q},\omega)
    V^{ext}_{0}(\mathbf{q},\omega) \;.
\end{equation}
This expression shows that even in the case of an external potential
with long wavelength variations in space, the response of the solid
will have shorter wavelength components that are commonly called local
field effects. The macroscopic dielectric constant $\epsilon_\text{M}$,
related to the measured optical properties in solids is the ratio
between the average of the total potential in one unit cell, i.e.
$V_{0}(\mathbf{q},\omega)$, and the external field. In this way we
identify the macroscopic dielectric constant 
\begin{equation}
  \label{eq:dielmacro}
    \epsilon_{\text{M}}(\mathbf{q},\omega)={1\over
    \bar{\epsilon}_{0,0}(\mathbf{q},\omega) }
\end{equation}
by the $(0,0)$ element of the {\em  inverse} dielectric tensor.

The expression for the macroscopic dielectric constant given in
Eq.~(\ref{eq:dielmacro}) is rather costly to evaluate. It involves the
evaluation of the dielectric tensor with components
$(\mathbf{G},\mathbf{G}^\prime)$ up to a certain cutoff and the
subsequent inversion to obtain the $(0,0)$ component of the inverse
tensor.  It is a common simplification to neglect the local field
effects and replace the $(0,0)$ component of the inverse by the
inverse of the $(0,0)$ component to obtain
\begin{equation}
  \label{eq:enlf}
  \epsilon_{\text{M}}^{\{\text{nlf}\}}(\mathbf{q},\omega) =
  \epsilon_{0,0}(\mathbf{q},\omega) =
  1 - v(\mathbf{q}) P^0_{0,0}(\mathbf{q},\omega) \;.
\end{equation}

We are going to neglect local field effects for the rest of the paper
in order to give the zero order theory evaluated with the Kohn-Sham
eigenvalues and eigenvectors from LAPW, one of the most precise
electronic structure methods available. Further improvements of the
theory can be introduced by including these local field effects or a
better many body description of the electron system. Our results
will be the starting point where no basis set effect can be blamed for
inaccuracies.  Early attempts to include local field corrections by
Stephen Adler \cite{Adler}, and independently but in the same spirit
by Nathan Wiser \cite{Wiser} found that, in a first approximation, these
effects can be described by the Lorenz-Lorentz formula with a
renormalized polarizability. The effect turned out to be quite small
for most Fermi surfaces studied. More recently, it has been found that
local field effects are important for high energy excitations where
localized states are involved.\cite{Vast}

The Lindhard expression for the dielectric constant of
Eq.~(\ref{eq:lindhard}) can be obtained from the expression of the
macroscopic dielectric constant without local field effects
(Eq.~(\ref{eq:enlf})) by assuming that the electron system is composed
by only one band and the matrix elements are equal to one. This is the
case for a simple parabolic band of free electrons.

Since we are interested in the optical properties of solids, and the
wavevector of light $\mathbf{q}$ is much smaller than any typical
wavevector of electrons in the system, we need to evaluate the
polarizability $P$ entering in Eq.~(\ref{eq:enlf}) in the limit of
$\mathbf{q}\to 0$. The matrix elements involved in the expression for
$P$ given in Eq.~(\ref{eq:p0q}) and defined in Eq.~(\ref{eq:matelem})
can be evaluated for small $\mathbf{q}$ by perturbation theory. We
will show the details in Appendix \ref{sec:mmee}, where the result is
given in Eq.~(\ref{eq:meqp}). This expression shows that the limit is
different for intraband matrix elements, i.e., those with $n'=n$, and
for interband matrix elements. According to this observation it is
convenient to split the sum over $n'$ and $n$ into those terms with
$n'=n$, corresponding to electronic transitions in the same band, and
those with $n' \neq n$ corresponding to interband transitions.  In this
manner, we can write
\begin{equation}
  \label{eq:interintra}
  \epsilon_M^{\{\text{nlf}\}}(\mathbf{q}\to 0,\omega) = 1 + 
  \epsilon^{\{\text{intra}\}}(\mathbf{q}\to 0,\omega) + 
  \epsilon^{\{\text{inter}\}}(\mathbf{q}\to 0,\omega) \;,
\end{equation}
where the intraband part of the dielectric constant is given by
\begin{equation}
  \label{eq:eintradef}
  \epsilon^{\{\text{intra}\}}(\mathbf{q}\to 0,\omega) =
-\lim_{\mathbf{q}\to 0}
  \frac{4\pi e^2}{\Omega_c\left|\mathbf{q}\right|^2}\;
  \sum_{n,\mathbf{k}}
  \frac{f_0(\varepsilon_{n,\mathbf{k}+\mathbf{q}}) - 
    f_0(\varepsilon_{n,\mathbf{k}})}{\varepsilon_{n,\mathbf{k}+\mathbf{q}} -
    \varepsilon_{n,\mathbf{k}}-\omega} \:
  \left|M_{n,n}^0(\mathbf{k},\mathbf{q})\right|^2 \; ,
\end{equation}
and the interband part by
\begin{equation}
  \label{eq:einterdef}
  \epsilon^{\{\text{inter}\}}(\mathbf{q}\to 0,\omega) =
-\lim_{\mathbf{q}\to 0}
  \frac{4\pi e^2}{\Omega_c\left|\mathbf{q}\right|^2}\;
  \sum_{n',n\neq n',\mathbf{k}}
  \frac{f_0(\varepsilon_{n',\mathbf{k}+\mathbf{q}}) - 
         f_0(\varepsilon_{n,\mathbf{k}})}
      {\varepsilon_{n',\mathbf{k}+\mathbf{q}}  - 
       \varepsilon_{n,\mathbf{k}}-\omega} \:
  \left|M_{n,n'}^0(\mathbf{k},\mathbf{q})\right|^2 \; .
\end{equation}

While taking the limit of $\mathbf{q}\to 0$ we can use the expression
of the matrix element evaluated in Eq.~(\ref{eq:meqp}) and the
expansion of the band energies obtained in Eq.~(\ref{eq:enqp}) to
obtain for the intraband part
\begin{equation}
  \label{eq:eintrafin}
  \epsilon^{\{\text{intra}\}}(\mathbf{q}\to 0,\omega) =
-\lim_{\mathbf{q}\to 0}
  \frac{4\pi \hbar^2 e^2}{\Omega_c m^2 \omega^2}\;
  \sum_{n,\mathbf{k}}
  \left(-\frac{\partial
      f}{\partial\varepsilon}\right)_{\varepsilon_{n,\mathbf{k}}} 
\left( \mathbf{p}_{n,n,\mathbf{k}} \cdot
  \frac{\mathbf{q}}{\left|\mathbf{q}\right|} \right)^2\;,
\end{equation}
where the derivative of the Fermi function with respect to the energy
should be considered as a restriction to sum only over the states at
the Fermi level. The momentum matrix element
$\mathbf{p}_{n,l,\mathbf{k}}$ is defined in Eq.(\ref{eq:mme}) and
evaluated for the LAPW basis set in Section \ref{sec:mme}. The
interband expression is
\begin{equation}
  \label{eq:einterfin}
  \epsilon^{\{\text{inter}\}}(\mathbf{q}\to 0,\omega) =
-\lim_{\mathbf{q}\to 0}
  \frac{4\pi \hbar^2 e^2}{\Omega_c m^2}\;
  \sum_\mathbf{k} \sum_{c,v}
  \frac{\left( \mathbf{p}_{c,v,\mathbf{k}} \cdot
  \mathbf{q}/\left|\mathbf{q}\right| \right)^2}
{
\left(\varepsilon_{c,\mathbf{k}}-\varepsilon_{v,\mathbf{k}}-\omega\right)
\left(\varepsilon_{c,\mathbf{k}}-\varepsilon_{v,\mathbf{k}}\right)^2}
\;,
\end{equation}
where, for a given $\mathbf{k}$, $c$ runs over the empty states and
$v$ over the occupied states.

From these expressions, we see that the limit is not well defined and
the result depends on the direction of the vector $\mathbf{q}$ even
when the limit is taken. This defines the dielectric constant for
$\mathbf{q}\to 0$ as a three dimensional tensor, the dielectric
tensor given by
\begin{eqnarray}
  \label{eq:dieltensq0}
  \epsilon_{i,j}(\omega) &=& \delta_{i,j}\\ \nonumber
&-&  
  \frac{4\pi \hbar^2 e^2}{\Omega_c m^2 \omega^2}\;
  \sum_{n,\mathbf{k}}
  \left(-\frac{\partial
      f}{\partial\varepsilon}\right)_{\varepsilon_{n,\mathbf{k}}} 
p_{i;n,n,\mathbf{k}}\:p_{j;n,n,\mathbf{k}}\\ \nonumber
&-& 
  \frac{4\pi \hbar^2 e^2}{\Omega_c m^2}\;
  \sum_\mathbf{k} \sum_{c,v}
  \frac{p_{i;c,v,\mathbf{k}}p_{j;c,v,\mathbf{k}}
}
{
\left(\varepsilon_{c,\mathbf{k}}-\varepsilon_{v,\mathbf{k}}-\omega\right)
\left(\varepsilon_{c,\mathbf{k}}-\varepsilon_{v,\mathbf{k}}\right)^2}
\;.
\end{eqnarray}

It is clear that in the derivation of the RPA formulas presented here
the unperturbed electronic states are described by the bare
one-particle propagator $G^0$. In the Kohn-Sham approximation the
many-body problem was reduced to the solution of N one-body problems.
These single-particle problems are not representing bare particles but
an effective decoupling through the introduction of independent
quasiparticles.  If the decoupling is done with the Hartree-Fock
approximation, the scheme we just presented is conserving in the
Kadanoff-Baym \cite{kadanoff-baym} sense.  Here, we are going to use
the RPA expression for the dielectric constant evaluated with the KS
orbitals.  This is equivalent to assume that the frequency dependent
exchange and correlation kernel of TDDFT is zero, which represents a
non conserving approximation to the problem. Nevertheless, in order to
gain physical insight into the applicability and limitations of this
procedure, we will use the RPA formula utilizing the KS orbitals as our
approximation to the dielectric constant. This procedure is also
called the independent particle approximation because it neglects the
electron-hole interaction during the absorption process. This should
be a good approximation for metals due to the more effective screening
of the Coulomb interaction compared to semiconductors.  Comparison
between our results and experimental data is an important step in the
validation of this approach. It also allows to compare, when possible,
with the results of more elaborated schemes to treat the many body
effects as GW, the soultion of the Bethe-Salpeter-Equation (BSE), or different frequency dependent kernels in TDDFT.

At the end of this Section, we give the expressions which will be
finally computed. Most important is the imaginary part of the interband contribution to the dielectric tensor components:
\begin{equation}
  \label{eq:eintercomp}
  \epsilon_{ij}^{\{\text{inter}\}}(\omega) =
  \frac{\hbar^2 e^2}{\pi  m^2 \omega^2}
  \sum_{n,n'} \int\limits_\mathbf{k}
  p_{i;n',n,\mathbf{k}}  p_{j;n',n,\mathbf{k}} \; 
  \left(   f({\varepsilon_{n,\mathbf{k}}}) -  
            f({\varepsilon_{n',\mathbf{k}}}) \right)
  \delta \left(
  \varepsilon_{n',\mathbf{k}}-\varepsilon_{n,\mathbf{k}} - \omega   
           \right)
\end{equation}
The corresponding real parts are obtained by Kramers-Kronig transformation.

As it is evident from Eq. (\ref{eq:dieltensq0}), the intraband part is singular at $\omega=0$. At this point the plasma frequency $\omega_{pl;ij}$ is defined by:
\begin{equation}
  \label{eq:omegapl}
  \omega_{pl;ij}^2 =
  \frac{\hbar^2 e^2}{\pi  m^2}
  \sum_{n} \int\limits_\mathbf{k}
  p_{i;n,n,\mathbf{k}}  p_{j;n,n,\mathbf{k}} \; 
   \delta \left(
  \varepsilon_{n,\mathbf{k}}-\varepsilon_{F} \right)
\end{equation}
In practical calculations, a lifetime broadening $\Gamma$ is introduced adopting a Drude-like shape for the intraband contribution.  $\epsilon^{\{\text{intra}\}}_{ij}(\omega)$ then reads:
\begin{eqnarray}
  \label{eq:edrude}
  {\mathrm Im} \epsilon_{ij}^{\{\text{intra}\}}(\omega) & = &
  \frac{\Gamma  \omega^2_{pl;ij}}{\omega (\omega^2+\Gamma^2)} \\
{\mathrm Re} \epsilon_{ij}^{\{\text{intra}\}}(\omega) & =&1 -
  \frac{ \omega^2_{pl;ij}}{(\omega^2+\Gamma^2)} \;
 \end{eqnarray}
Note that the imaginary part is still singular; therefore one usually works with the optical conductivity. The BZ integrations in Eqs. (\ref{eq:eintercomp}) and (\ref{eq:omegapl}) are carried out by the linear tetrahedron method.
\subsection{Symmetry}

The dielectric tensor is symmetric with up to
six independent components according to the symmetry of the
crystal. Therefore the general expression for the imaginary part of
$\epsilon$ is:
\begin{eqnarray}
\left(
\begin{array}{ccc}
\mathrm{Im}\:\epsilon_{xx} & \mathrm{Im}\:\epsilon_{xy} & \mathrm{Im}\:\epsilon_{xz}\\
\mathrm{Im}\:\epsilon_{xy} & \mathrm{Im}\:\epsilon_{yy} & \mathrm{Im}\:\epsilon_{yz}\\
\mathrm{Im}\:\epsilon_{xz} & \mathrm{Im}\:\epsilon_{yz} & \mathrm{Im}\:\epsilon_{zz}
\end{array}
\right)
\end{eqnarray}
For orthorhombic or higher symmetry only diagonal components exist. In
case of cubic symmetry the optical properties are isotropic, i.e.,
there is only one independent component $\mathrm{Im}\:\epsilon_{xx}$,
while for uniaxial symmetry (tetragonal, hexagonal) and orthorhombic
symmetry there are two and three independent components, respectively.
In the monoclinic case non-diagonal elements occur, i.e for $\gamma
\ne 90^\circ$ the tensor takes the form:
\begin{eqnarray}
\left (
\begin{array}{ccc}
\mathrm{Im}\:\epsilon_{xx} & \mathrm{Im}\:\epsilon_{xy} & 0 \\
\mathrm{Im}\:\epsilon_{xy} & \mathrm{Im}\:\epsilon_{yy} & 0 \\
0  & 0  & \mathrm{Im}\:\epsilon_{zz}
\end{array}
\right )
\end{eqnarray}
These materials are optically active or birefringent. Only in the
triclinic crystal all six components are different.

These symmetry considerations only hold for non-relativistic and
scalar-relativistic calculations. In these cases the spin-up and
spin-down elements can be evaluated separately, where the
corresponding real parts are obtained by the Kramers--Kronig
transformation (see below). In case of magento-optics the symmetry is
reduced by the presence of a magnetic field as well as by the loss of
time-reversal symmetry due to spin-orbit coupling. The latter gives
rise to antisymmetric non-diagonal components, e.g. for the magnetic
field parallel to $z$:
\begin{eqnarray}
\left(
\begin{array}{ccc}
0 & \mathrm{Re}\:\epsilon_{xy} & 0 \\
-\mathrm{Re}\:\epsilon_{xy} & 0 & 0 \\
0  & 0  & 0
\end{array}
\right).
\end{eqnarray}
The corresponding imaginary parts are again obtained by
Kramers--Kronig analysis. In case of monoclinic or lower symmetry the
off-diagonal elements can therefore have two independent
contributions, one due to a non-orthogonal crystal axis, and the other
one due to the magneto-optical effect.
 
\subsection{Kramers--Kronig relations and optical constants}

From the imaginary part of the dielectric tensor component
$\mathrm{Im}\:\epsilon_{ij}$ the corresponding real part is
obtained by
\begin{equation}
\mathrm{Re}\: \epsilon_{ij} = \delta_{ij} + \frac{2}{\pi} \wp
\int_0^{\infty}
\frac{\omega'\:\mathrm{Im}\:\epsilon_{ij}(\omega')}{\omega'^2-\omega^2} d\omega'.
\end{equation}
Given the real part, the inverse transformation has to be used.

With the knowledge of the complex dielectric tensor components all
other frequency-dependent optical "constants" can be obtained. The
most often used ones are the real part of the optical conductivity
\begin{equation}
\mathrm{Re}\: \sigma_{ij} (\omega)= \frac{\omega}{4\pi} \mathrm{Im}\:\epsilon_{ij}(\omega),
\end{equation}
the loss function 
\begin{equation}
L_{ij}(\omega) = -
\mathrm{Im}\: \left( \frac{1}{\epsilon_{ij}(\omega)} \right),
\end{equation}
and the reflectivity at normal incidence
\begin{equation}
R_{ii}(\omega) = \frac{(\mathrm{n}-1)^2+\mathrm{k}^2}{(\mathrm{n}+1)^2+\mathrm{k}^2}
\end{equation}
with $\mathrm{n}$ and $\mathrm{k}$ being the real in imaginary part of
the complex refractive index (refractive index and extinction
coefficient):
\begin{equation}
\mathrm{n_{ii}(\omega)} = \sqrt{ \frac{
\left |\epsilon_{ii}(\omega) \right |
+ \mathrm{Re}\:\epsilon_{ii}(\omega)} {2}} 
\end{equation}
\begin{equation}
\mathrm{k_{ii}(\omega)} = \sqrt{ \frac{ 
\left |\epsilon_{ii}(\omega) \right |
- \mathrm{Re}\:\epsilon_{ii}(\omega)} {2}} 
\end{equation}
The absorption coefficient is given by
\begin{equation}
A_{ii}(\omega) = \frac{2\omega \: \mathrm{k(\omega)} }{c}
.
\end{equation}
Note that the latter formulas only hold for the diagonal form of the
tensor.

\subsection{Sumrules}
There are three sumrules which obtain information about the absorption process:
\begin{equation}
\int\limits_0^{\omega^{\prime}}
\sigma \left( \omega \right) \omega \:d\omega
= N_{\mathit{eff}} \left( \omega^{\prime} \right)
\label{sumrule1}
\end{equation}
\begin{equation}
\int\limits_0^{\omega^{\prime}}
\mathrm{Im}\: \left( \frac{-1} {\epsilon \left( \omega \right)} \right) 
\omega \:d\omega
 = N_{\mathit{eff}} \left( \omega^{\prime} \right)
\label{sumrule2}
\end{equation}
\begin{equation}
\int\limits_0^{\infty}
\mathrm{Im}\: \left( \frac{-1} {\epsilon \left( \omega \right)} \right) 
\frac{1} {\omega} \:d\omega
 = \frac{\pi}{2}
\label{sumrule3}
\end{equation}
The first two give an effective number of electrons contributing to
the absorption process as a function of energy. Typically, in the low
energy region the contribution to the intraband spectrum should sum up
to the number of the outermost valence electrons.

\section{Optical response within the LAPW method}
\label{optic_LAPW}

\subsection{The LAPW basis set \label{basis_LAPW}}
In bandstructure calculations based on density-functional theory
\cite{Kohn:64} the single-particle electronic states 
$\Psi _{n {\bf k}}({\bf r})$ and energies $\varepsilon_{n{\bf k}}$ 
are described by the solutions of the Kohn-Sham (KS) equation \cite{Sham:65}
\begin{equation}
\left[ -\frac{\hbar^2}{2m}\nabla^{2} + V_{\mathit{eff}}({\bf r}) \right] 
\Psi_{n {\bf k}}({\bf r}) = \varepsilon_{n{\bf k}} \Psi_{n {\bf k}}({\bf r})
\label{KS-equation}
\end{equation}
with the effective potential $V_{\mathit{eff}}({\bf r})$ being the sum of the bare
Coulomb potential of the atomic nuclei $V_{latt}({\bf r})$, the Hartree
potential $V_{H}({\bf r})$ and the exchange correlation potential 
$V_{xc}({\bf r})$. In practical calculations, Eq.\ (\ref{KS-equation}) is
solved via the Rayleigh-Ritz variational principle. In this procedure the
electronic states $\Psi _{n {\bf k}}({\bf r})$ (KS-orbitals) are first 
expanded in terms of a physically appropriate finite
set of basis functions $\left\{ \phi _{{\bf {k+G}}}\right\} $, 
\begin{equation}
\Psi_{n {\bf k}}({\bf r}) =
\sum_{{\bf G}} C_{n{\bf k}}({\bf G}) \phi_{{\bf {k+G}}}({\bf r})  
\label{ansatz}
\end{equation}
with ${\bf G}$ and $C_{n{\bf k}}({\bf G})$ denoting a reciprocal
lattice vector, and the corresponding variational coefficient, respectively. To
determine $C_{n{\bf k}}({\bf G})$ the ansatz (\ref{ansatz}) is
inserted into Eq.\ (\ref{KS-equation}) followed by the minimization of the
total crystal energy with respect to the variational coefficients. 
The eigenvectors and eigenvalues of the resulting matrix equation, 
\begin{equation}
\sum_{{\bf G}^{\prime}}
(H_{{\bf {k+G}},{\bf {k+G}}^{\prime}} -
\varepsilon_{n {\bf k}} S_{{\bf {k+G}},{\bf {k+G}}^{\prime}})
C_{n {\bf k}}({\bf G}^{\prime})=0 
\text{ \ \ \ \ } {\bf k} \in \text{BZ}  
\label{matrix_equation}
\end{equation}
with 
\begin{equation}
H_{{\bf {k+G}},{\bf {k+G}}^{\prime}} \equiv 
\langle \phi_{{\bf {k+G}}}                
               \left| - \frac{\hbar^2}{2m}\nabla^{2} + V_{\mathit{eff}} \right| 
        \phi_{{\bf {k+G}}^{\prime}} 
\rangle_{\Omega_c}
\label{H_matrix}
\end{equation}
\begin{equation}
S_{{\bf {k+G}},{\bf {k+G}}^{\prime}} \equiv 
\langle \phi_{{\bf {k+G}}} 
               \left| \phi_{{\bf {k+G}}^{\prime}} \right  
\rangle_{\Omega_c}
\label{S_matrix}
\end{equation}
being Hamilton and overlap matrix, respectively, finally provide the
numerical values for $C_{n{\bf k}}({\bf G})$ and $\varepsilon_{n {\bf k}}$. 
$\Omega_c$ is the volume of the unit cell which in the LAPW method
\cite{Andersen:75,Singh:94} is partitioned into an interstitial region ($\mathrm{I}$) 
and non-overlapping muffin-tin spheres ($\mathrm{MT}_{\alpha }$) centered on the atomic 
nuclei. The corresponding basis functions are defined as, 
\begin{equation}
\phi_{{\bf {k+G}}}({\bf r}) = 
\frac{1}{\sqrt{\Omega_c}}e^{i({\bf {k+G}}) {\bf r}},
\text{ } {\bf r} \in \mathrm{I}  
\label{basis_Int}
\end{equation}
and 
%
\begin{eqnarray}
\phi_{{\bf {k+G}}}  ({\bf S}_{\alpha} + {\bf r}) & = &
\sum_{lm}  
\underbrace{
\left[
    A_{lm}^{\alpha}({\bf {k+G}})       u_{l}^{\alpha}(r,E_l) +
    B_{lm}^{\alpha}({\bf {k+G}}) \dot{u}_{l}^{\alpha}(r,E_l) \right] 
}_{ W_{lm}^{ \alpha,{\bf {k+G}} } } Y_{l,m}( {\bf {{\hat r}}} )   \\           
& = &  
\sum_{lm} 
        W_{lm}^{ \alpha,{\bf {k+G}} } (r,E_l) Y_{l,m} ({\bf{{\hat r}}})
\text{ \ \ } \left| {\bf r} \right| \leq R_{\alpha},  
\label{basis_MT}
\end{eqnarray}
%
where ${\bf S}_{\alpha}$ denotes the position vector of the atomic
nucleus $\alpha$.  The radius of the corresponding muffin-tin sphere
is $R_{\alpha }$.  The product of the spherical harmonic
$Y_{l,m}(\bf{\hat{r}})$ and the radial function $u_{l}^{\alpha }(r)$
is the solution of the Schr\"{o}dinger equation for a spherical
symmetric potential where the eigenvalue has been replaced by an
appropriate energy parameter. The second radial function
$\dot{u}_{l}^{\alpha }(r)$ is the derivative of the first one with
respect to the energy. The coefficients
$A_{lm}^{\alpha }({\bf {k+G}})$ and $B_{lm}^{\alpha }({\bf {k+G}})$ 
are determined for each atom by imposing the requirements that the
basis function has to be continuous in value and slope on the
MT-surfaces:

\begin{eqnarray}
A_{lm}^{\alpha }({\bf {k+G}}) & = & \frac{4\pi}{\sqrt{\Omega}} i^l
Y_{lm}^{\ast} \left( \widehat{\bf {k+G}}\right) R_{\alpha}^2\: e^{i {\bf (k+G)
S_{\alpha}}}
a_l({\bf {k+G}})\\
B_{lm}^{\alpha }({\bf {k+G}}) & = &\frac{4\pi}{\sqrt{\Omega}} i^l
Y_{lm}^{\ast} \left( \widehat{\bf {k+G}}\right) R_{\alpha}^2\: e^{i {\bf (k+G) S_{\alpha}}}
b_l({\bf {k+G}})
\end{eqnarray}
with
\begin{eqnarray}
a_{l} = j_l^{\prime} \left( |{\bf k+G}| R_{\alpha}\right) \dot{u}_l(R_{\alpha})
- j_l \left( |{\bf k+G}| R_{\alpha}\right) \dot{u}_l^{\prime} (R_{\alpha}) \\
b_{l} = j_l \left( |{\bf k+G}| R_{\alpha}\right) {u}_l(R_{\alpha})
- j_l^{\prime} \left( |{\bf k+G}| R_{\alpha}\right) {u}_l^{\prime} (R_{\alpha}) 
\end{eqnarray}

For the explicit evaluation of the Hamilton matrix elements (\ref{H_matrix}) 
an appropriate dual representation of the effective potential is needed. 
Within the atomic spheres the potential is expanded into spherical harmonics 
and in the interstitial it is represented by a Fourier series.

\subsection{The momentum matrix elements}
\label{sec:mme}
Due to the dual representation of the LAPW basis functions, the momentum matrix
element is a sum of contributions from the atomic spheres as well as from the
interstitial region:
\begin{equation}
\langle n^{\prime}{\bf k} \left| \mathbf{p} \right| n{\bf k} \rangle = \sum_\alpha 
\langle n^{\prime}{\bf k} \left| \mathbf{p} \right| n{\bf k} \rangle_{\mathrm{MT}_{\alpha}} +
\langle n^{\prime}{\bf k} \left| \mathbf{p} \right| n{\bf k} \rangle_{\mathrm{I}}
\label{ME_sum}
\end{equation}
\subsubsection{Contributions from the atomic spheres}
\label {ME_spheres}
The usage of spherical harmonics in the LAPW basis set suggest to 
calculate the expressions
$ \langle n^{\prime}{\bf k} \left| \partial x + i \partial y \right| n{\bf k} \rangle$
and 
$\langle n^{\prime}{\bf k} \left| \partial x - i \partial y) \right| n{\bf k} \rangle$
and to derive the $x-$ and $y-$ component as their linear combinations.
Taking into account the ansatz (\ref{ansatz}) for the 
wavefunctions the following expressions have to be evaluated:
\begin{eqnarray}
^{\alpha}\Phi_{{\bf {k+G^{\prime}},\bf {k+G}}}^{x+iy} & \equiv &
\langle \phi_{\bf {k+G^{\prime}}} ( {\bf S}_{\alpha} + {\bf r} )
\left| \partial x + i\partial y \right| 
\phi_{\bf {k+G}} ( {\bf S}_{\alpha} + {\bf r} ) \rangle \label{Phi_x+y}\\
^{\alpha}\Phi_{{\bf {k+G^{\prime}},\bf {k+G}}}^{x-iy} & \equiv &
\langle \phi_{\bf {k+G^{\prime}}} ( {\bf S}_{\alpha} + {\bf r} )
\left| \partial x - i\partial y \right| 
\phi_{\bf {k+G}} ( {\bf S}_{\alpha} + {\bf r} ) \rangle \label{Phi_x-y}\\
^{\alpha}\Phi_{{\bf {k+G^{\prime}},\bf {k+G}}}^{z} & \equiv &
\langle \phi_{\bf {k+G^{\prime}}} ( {\bf S}_{\alpha} + {\bf r} )
\left| \partial z \right| 
\phi_{\bf {k+G}} ( {\bf S}_{\alpha} + {\bf r} ) \rangle
\label{Phi_z}
\end{eqnarray}
Expressing $\partial x + i\partial y$, $\partial x - i\partial y$, and $\partial z$ 
in spherical coordinates 
\begin{eqnarray}
\partial x \pm i\partial y & = & \sin \theta e^ {\pm i\phi} 
\frac {\partial} {\partial r} + \frac {1} {r}  e^ {\pm i\phi} 
\left ( \cos \theta \frac {\partial} {\partial \theta} 
\pm \frac {i} {\sin \theta} \frac {\partial} {\partial \phi} \right ) \\
\partial z & = & \cos \theta \frac {\partial} {\partial r} - 
\frac {1} {r} \sin \theta \frac {\partial} {\partial \theta}
\label{partials}
\end{eqnarray}
we can elaborate the matrix elements (\ref{Phi_x+y}--\ref{Phi_z}). We explicitly demonstrate
the procedure for the first component:
\begin{equation}
^{\alpha}\Phi_{{\bf {k+G^{\prime}},\bf {k+G}}}^{x+iy}  =
\langle  \phi_{\bf {k+G^{\prime}}} ( {\bf S}_{\alpha} + {\bf r} ) \left| 
\sin \theta e^ {i\phi} \frac {\partial} {\partial r} + 
\frac {1}{r}  e^ {i\phi}  \left ( \cos \theta \frac {\partial} {\partial \theta} +
\frac {i} {\sin \theta} \frac {\partial} {\partial \phi} \right ) 
\right| \phi_{\bf {k+G}} ( {\bf S}_{\alpha} + {\bf r} ) \rangle
\label{Phi_x+iy}
\end{equation}
Here the first term of the operator acts on the radial coordinate only, whereas
the second term acts on the angular coordinates only. Applying this operator to 
the LAPW basis functions (\ref{basis_MT}) one obtains:
%
\begin{eqnarray}
\left ( \partial x + i \partial y \right ) 
\phi_{ {\bf {k+G}} } ( {\bf S}_{\alpha} + {\bf r} ) = 
& &\sum_{lm} 
  \frac{\partial} {\partial r} W_{lm}^{\alpha,{\bf {k+G}} } (r) 
  \sin \theta  e^ { i\phi} Y_{l,m} ( {\bf {{\hat r}} } ) +  \nonumber \\
\frac{1}{r} & &\sum_{lm}  
W_{lm}^{\alpha,{\bf {k+G}}} (r)
e^ { i\phi} \left ( \cos \theta \frac {\partial} {\partial \theta} +
\frac {i} {\sin \theta} \frac {\partial} {\partial \phi} \right )
Y_{l,m} ( {\bf {\hat r} } ) 
\label{d_phi}
\end{eqnarray}
%
Exploiting the relations between the spherical harmonics as summarized in the
Appendix, and omitting the arguments of the spherical harmonics for simplicity, 
Eq. (\ref{d_phi}) becomes
%
\begin{eqnarray}
\left ( \partial x + i \partial y \right ) 
\phi_{{\bf {k+G}}} ({\bf S}_{\alpha} + {\bf r}) = 
\sum_{lm} 
& & \left[ \frac{\partial} {\partial r} W_{lm}^{\alpha,{\bf {k+G}} } (r)
                             - \frac{l}{r}    W_{lm}(r) \right] F_{l,m}^{(1)} Y_{l+1,m+1} +  
\nonumber \\
& & \left[\frac{\partial} {\partial r} W_{lm}^{\alpha,{\bf {k+G}} } (r)
                            + \frac{l+1}{r}  W_{lm}(r) \right] F_{l,m}^{(2)} Y_{l-1,m+1}\;.
\label{d_phi_short}
\end{eqnarray}
%
The $x+iy$ component will then be
\begin{eqnarray}
& &^{\alpha}\Phi_{{\bf {k+G^{\prime}},\bf {k+G}}}^{x+iy} = 
\int_{\mathrm{MT}_{\alpha}} 
\phi_{{\bf {k+G^{\prime}}}} ({\bf S}_{\alpha} + {\bf r})
\left[ \partial x + i\partial y \right] 
\phi_{{\bf {k+G}}} ({\bf S}_{\alpha} + {\bf r}) d{\bf r} =  \nonumber \\
& &\int\limits_0^{R_{\alpha}} r^2 dr \oint d\Omega 
\sum_{l^{\prime}m^{\prime}} \sum_{lm} 
W_{l^{\prime}m^{\prime}}^{\ast \alpha,{\bf {k+G^{\prime}}}} 
Y_{l^{\prime},m^{\prime}}^{\ast} \times \nonumber \\ 
& & \left ( 
\left[ \frac {\partial} {\partial r} W_{lm}^{\alpha,{\bf {k+G}} } (r)
                              - \frac{l}{r}  W_{lm}(r) \right] F_{l,m}^{(1)} Y_{l+1,m+1} +  
\left[ \frac {\partial} {\partial r} W_{lm}^{\alpha,{\bf {k+G}} } (r)
                              + \frac{l+1}{r} W_{lm}(r) \right] F_{l,m}^{(2)} Y_{l-1,m+1}
\right ) = \nonumber \\
& &\sum_{ l^{\prime}m^{\prime} } \sum_{lm}
\underbrace{
\int\limits_0^{R_{\alpha}} r^2 dr 
W_{l^{\prime}m^{\prime}}^{\ast \alpha,{\bf {k+G^{\prime}}}} 
\left [ \frac {\partial} {\partial r} W_{lm}^{\alpha,{\bf {k+G}} } (r)
                              - \frac{l}{r}   W_{lm}(r) \right ] F_{l,m}^{(1)}
}_{R_{l,m}^{l^{\prime},m^{\prime}}(r)}
\underbrace{
\oint d\Omega Y_{l^{\prime},m^{\prime}}^{\ast}  Y_{l+1,m+1} 
}_{{\delta_{l^{\prime},l+1}}{\delta_{m^{\prime},m+1}}} + \nonumber \\ 
& & 
\sum_{l^{\prime}m^{\prime}} \sum_{lm}
\underbrace{
\int\limits_0^{R_{\alpha}} r^2 dr 
W_{l^{\prime}m^{\prime}}^{\ast \alpha,{\bf {k+G^{\prime}}}} 
\left[ \frac {\partial} {\partial r} W_{lm}^{\alpha,{\bf {k+G}} } (r)
                              + \frac{l+1}{r} W_{lm}(r) \right] F_{l,m}^{(2)} 
}_{T_{l,m}^{l^{\prime},m^{\prime}} (r) }
\underbrace{
\oint d\Omega Y_{l^{\prime},m^{\prime}}^{\ast} Y_{l-1,m+1}
}_{{\delta_{l^{\prime},l-1}}{\delta_{m^{\prime},m+1}}}.
\label{ME_basis_x+iy}
\end{eqnarray}
Writing explicitly the sums over $l$, $m$, $l^{\prime}$, and $m^{\prime}$, taking into 
account an upper limit $l_{max}$ for $l$ we obtain:
\begin{eqnarray}
^{\alpha}\Phi_{{\bf {k+G^{\prime}},\bf {k+G}}}^{x+iy} & = &
\sum_{l^{\prime}=0}^{lmax} \sum_{m^{\prime}=-l^{\prime}}^{l^{\prime}}
\sum_{l=0}^{lmax} \sum_{m=-l}^{l} 
R_{l,m}^{l^{\prime},m^{\prime}} (r)
\delta_{l^{\prime},l+1} \delta_{m^{\prime},m+1} + \nonumber \\ & &
\sum_{l^{\prime}=0}^{lmax} \sum_{m^{\prime}=-l^{\prime}}^{l^{\prime}}
\sum_{l=0}^{lmax} \sum_{m=-l}^{l} 
T_{l,m}^{l^{\prime},m^{\prime}} (r)
\delta_{l^{\prime},l-1} \delta_{m^{\prime},m+1} = \nonumber \\ & &
\sum_{l=0}^{lmax-1} \sum_{m=-l}^{l} \left( R_{l,m}^{l+1,m+1} (r) + 
T_{l+1,m-1}^{l,m} (r) \right)
\label{sum_x+iy}
\end{eqnarray}
Evaluating $R_{l,m}^{l+1,m+1}(r)$ and $T_{l+1,m-1}^{l,m}(r)$ using Eqs.
(\ref{ME_basis_x+iy}) and (\ref{basis_MT}) we finally obtain:
\newpage
\begin{eqnarray}
\begin{array}{lll}
^{\alpha}\Phi_{{\bf {k+G^{\prime}},\bf {k+G}}}^{x+iy} = 
& \sum\limits_{l=0}^{lmax-1} \sum\limits_{m=-l}^{l} 
\left \{ F_{lm}^{(1)} \right.  \\ 
& \left \{ 
A_{l+1,m+1}^{^\ast \alpha} ( {\bf {k+G^ { \prime} } } ) 
A_{l,m}^{\alpha}          ( {\bf {k+G} } )
\left ( \int\limits_0^{R_{\alpha}} u_{l+1}(r) u_{l}^{\prime}(r) r^2 dr -
      l \int\limits_0^{R_{\alpha}} u_{l+1}(r) u_{l}         (r) r   dr        \right )
\right.  + \\ 
&   
A_{l+1,m+1}^{^\ast \alpha} ( {\bf {k+G^{\prime} } })  
B_{l,m}^{\alpha}          ( {\bf {k+G} } ) 
\left ( \int\limits_0^{R_{\alpha}} u_{l+1}(r) \dot{u}_{l}^{\prime}(r) r^2 dr -
      l \int\limits_0^{R_{\alpha}} u_{l+1}(r) \dot{u}_{l}           (r) r   dr \right ) 
+  \\
& 
B_{l+1,m+1}^{^\ast \alpha} ( {\bf {k+G^ {\prime} } } ) 
A_{l,m}^{\alpha}          ( {\bf {k+G} } ) 
\left ( \int\limits_0^{R_{\alpha}} \dot{u}_{l+1}(r) u_{l}^{\prime}(r) r^2 dr -
      l \int\limits_0^{R_{\alpha}} \dot{u}_{l+1}(r) u_{l}         (r) r   dr  \right )
+  \\ 
& \left.
B_{l+1,m+1}^{^\ast \alpha} ( {\bf {k+G^ {\prime} } } ) 
B_{l,m}^{\alpha}          ( {\bf {k+G} } )
\left ( \int\limits_0^{R_{\alpha}} \dot{u}_{l+1}(r) \dot{u}_{l}^{\prime}(r) r^2 dr -
      l \int\limits_0^{R_{\alpha}} \dot{u}_{l+1}(r) \dot{u}_{l}         (r) r   dr  \right ) 
\right \}  \\ 
& +  F_{l+1,m-1}^{(2)}  \\ 
& \left \{ 
A_{l,m}^{^\ast \alpha} ( {\bf {k+G^{\prime} } } ) 
A_{l+1,m-1}^{\alpha}  ( {\bf {k+G} } ) 
\left( \int\limits_0^{R_{\alpha}} u_{l}(r) u^{\prime}_{l+1}(r) r^2 dr +
 (l+2) \int\limits_0^{R_{\alpha}} u_{l}(r) u_{l+1}(r) r dr \right) 
+ \right.  \\ 
& 
A_{l,m}^{^\ast \alpha} ( {\bf {k+G^{\prime} } } ) 
B_{l+1,m-1}^{\alpha}  ( {\bf {k+G} } ) 
\left( \int\limits_0^{R_{\alpha}}  u_{l}(r) \dot{u}^{\prime}_{l+1}(r) r^2 dr +
  (l+2) \int\limits_0^{R_{\alpha}} u_{l}(r) \dot{u}_{l+1}(r) r dr \right) 
+ \\ 
& 
B_{l,m}^{^\ast \alpha} ( {\bf {k+G ^{\prime} } } ) 
A_{l+1,m-1}^{\alpha}  ( {\bf {k+G} } ) 
\left( \int\limits_0^{R_{\alpha}} \dot{u}_{l}(r) u^{\prime}_{l+1}(r) r^2 dr +
(l+2) \int\limits_0^{R_{\alpha}} \dot{u}_{l}(r) u_{l+1}(r) r dr \right) 
+ \\ 
& \left. \left. 
B_{l,m}^{^\ast \alpha} ( {\bf {k+G ^{\prime} } } ) 
B_{l+1,m-1}^{\alpha}  ( {\bf {k+G } } )
\left( \int\limits_0^{R_{\alpha}} \dot{u}_{l}(r) \dot{u}^{\prime}_{l+1}(r) r^2 dr +
 (l+2) \int\limits_0^{R_{\alpha}} \dot{u}_{l}(r) \dot{u}_{l+1}(r) r dr \right)  
\right \}
\right \}\\
\end{array}
\nonumber\\
\label{Phi_x+iy_final}
\end{eqnarray}
%
with $u^{\prime}_{l}(r)$ denoting $\partial u_{l}(r)$/$\partial r$.
Analogously we derive the other two components:
\begin{eqnarray}
\begin{array}{lll}
^{\alpha}\Phi_{{\bf {k+G^{\prime}},\bf {k+G}}}^{x-iy} = 
& \sum\limits_{l=0}^{lmax-1} \sum\limits_{m=-l}^{l} 
  \left \{ F_{lm}^{(3)} \right.  \\ 
& \left \{  
A_{l-1,m+1}^{^\ast \alpha}({\bf {k+G^{\prime}}}) 
A_{l,m}^{\alpha}({\bf {k+G}}) 
\left ( \int\limits_0^{R_{\alpha}} u_{l+1}(r) u_{l}^{\prime}(r) r^2 dr -
     l \int\limits_0^{R_{\alpha}} u_{l+1}(r) u_{l}         (r) r   dr          \right )
+ \right.  \\ 
& 
A_{l-1,m+1}^{^\ast \alpha}({\bf {k+G^{\prime}}}) 
B_{l,m}^{\alpha}({\bf {k+G}}) 
\left ( \int\limits_0^{R_{\alpha}} u_{l+1}(r) \dot{u}_{l}^{\prime}(r) r^2 dr -
      l \int\limits_0^{R_{\alpha}} u_{l+1}(r) \dot{u}_{l}           (r) r   dr  \right )
+  \\
& 
B_{l-1,m+1}^{^\ast \alpha}({\bf {k+G^{\prime}}}) 
A_{l,m}^{\alpha}({\bf {k+G}}) 
\left ( \int\limits_0^{R_{\alpha}} \dot{u}_{l+1}(r) u_{l}^{\prime}(r) r^2 dr -
      l \int\limits_0^{R_{\alpha}} \dot{u}_{l+1}(r) u_{l}         (r) r   dr    \right )
+  \\ 
& \left.
B_{l-1,m+1}^{^\ast \alpha}({\bf {k+G^{\prime}}}) 
B_{l,m}^{\alpha}({\bf {k+G}})
\left ( \int\limits_0^{R_{\alpha}} \dot{u}_{l+1}(r) \dot{u}_{l}^{\prime}(r) r^2 dr -
      l \int\limits_0^{R_{\alpha}} \dot{u}_{l+1}(r) \dot{u}_{l}         (r) r   dr  \right ) 
\right \}  \\ 
& + F_{l+1,m-1}^{(4)}  \\ 
& \left \{ 
A_{l,m}^{^\ast \alpha}({\bf {k+G^{\prime}}}) 
A_{l+1,m+1}^{\alpha}({\bf {k+G}}) 
\left( \int\limits_0^{R_{\alpha}} u_{l}(r) u^{\prime}_{l+1}(r) r^2 dr +
 (l+2) \int\limits_0^{R_{\alpha}} u_{l}(r) u_{l+1}(r) r dr \right) + \right.  \\ 
& 
A_{l,m}^{^\ast \alpha}({\bf {k+G^{\prime}}}) 
B_{l+1,m+1}^{\alpha}({\bf {k+G}}) 
\left( \int\limits_0^{R_{\alpha}}  u_{l}(r) \dot{u}^{\prime}_{l+1}(r) r^2 dr +
  (l+2) \int\limits_0^{R_{\alpha}} u_{l}(r) \dot{u}_{l+1}(r) r dr \right) +  \\ 
&  
B_{l,m}^{^\ast \alpha}({\bf {k+G^{\prime}}}) 
A_{l+1,m+1}^{\alpha}({\bf {k+G}}) 
\left( \int\limits_0^{R_{\alpha}} \dot{u}_{l}(r) u^{\prime}_{l+1}(r) r^2 dr +
(l+2) \int\limits_0^{R_{\alpha}} \dot{u}_{l}(r) u_{l+1}(r) r dr \right) +  \\ 
& \left. \left. 
B_{l,m}^{^\ast \alpha}({\bf {k+G^{\prime}}}) 
B_{l+1,m+1}^{\alpha}({\bf {k+G}})
\left( \int\limits_0^{R_{\alpha}} \dot{u}_{l}(r) \dot{u}^{\prime}_{l+1}(r) r^2 dr 
 (l+2) \int\limits_0^{R_{\alpha}} \dot{u}_{l}(r) \dot{u}_{l+1}(r) r dr \right) 
\right \}
\right \} \\
\end{array}
\nonumber \\
\label{Phi_x-iy_final}
\end{eqnarray}
\begin{eqnarray}
\label{Phi_z_final}
\begin{array}{lll}
^{\alpha}\Phi_{{\bf {k+G^{\prime}},\bf {k+G}}}^{z} = 
& \sum\limits_{l=0}^{lmax-1} \sum\limits_{m=-l}^{l} 
\left \{ F_{lm}^{(5)}  \right.  \\ 
&  \left \{
A_{l+1,m+1}^{^\ast \alpha}({\bf {k+G^{\prime}}}) 
A_{l,m}^{\alpha}({\bf {k+G}}) 
\left ( \int\limits_0^{R_{\alpha}} u_{l+1}(r) u_{l}^{\prime}(r) r^2 dr -
     l \int\limits_0^{R_{\alpha}} u_{l+1}(r) u_{l}         (r) r   dr          \right )
+ \right.  \\ 
& A_{l+1,m+1}^{^\ast \alpha}({\bf {k+G^{\prime}}}) 
  B_{l,m}^{\alpha}({\bf {k+G}}) 
\left ( \int\limits_0^{R_{\alpha}} u_{l+1}(r) \dot{u}_{l}^{\prime}(r) r^2 dr -
      l \int\limits_0^{R_{\alpha}} u_{l+1}(r) \dot{u}_{l}           (r) r   dr  \right
) +
 \\
& B_{l+1,m+1}^{^\ast \alpha}({\bf {k+G^{\prime}}}) 
  A_{l,m}^{\alpha}({\bf {k+G}}) 
\left ( \int\limits_0^{R_{\alpha}} \dot{u}_{l+1}(r) u_{l}^{\prime}(r) r^2 dr -
     l \int\limits_0^{R_{\alpha}} \dot{u}_{l+1}(r) u_{l}         (r) r   dr    \right )
+ 
 \\ 
& \left. 
B_{l+1,m+1}^{^\ast \alpha}({\bf {k+G^{\prime}}}) 
B_{l,m}^{\alpha}({\bf {k+G}})
\left ( \int\limits_0^{R_{\alpha}} \dot{u}_{l+1}(r) \dot{u}_{l}^{\prime}(r) r^2 dr -
      l \int\limits_0^{R_{\alpha}} \dot{u}_{l+1}(r) \dot{u}_{l}         (r) r   dr  \right ) 
\right \}  \\ 
& +  F_{l+1,m}^{(6)}   \\ 
&  \left \{ 
A_{l,m}^{^\ast \alpha}({\bf {k+G^{\prime}}}) 
A_{l+1,m+1}^{\alpha}({\bf {k+G}}) 
\left( \int\limits_0^{R_{\alpha}} u_{l}(r) u^{\prime}_{l+1}(r) r^2 dr +
 (l+2) \int\limits_0^{R_{\alpha}} u_{l}(r) u_{l+1}(r) r dr \right) +
\right.  \\ 
& 
A_{l,m}^{^\ast \alpha}({\bf {k+G^{\prime}}}) 
B_{l+1,m+1}^{\alpha}({\bf {k+G}}) 
\left( \int\limits_0^{R_{\alpha}}  u_{l}(r) \dot{u}^{\prime}_{l+1}(r) r^2 dr +
 (l+2) \int\limits_0^{R_{\alpha}} u_{l}(r) \dot{u}_{l+1}(r) r dr \right) +  \\ 
& 
B_{l,m}^{^\ast \alpha}({\bf {k+G^{\prime}}}) 
A_{l+1,m+1}^{\alpha}({\bf {k+G}}) 
\left( \int\limits_0^{R_{\alpha}} \dot{u}_{l}(r) u^{\prime}_{l+1}(r) r^2 dr +
 (l+2) \int\limits_0^{R_{\alpha}} \dot{u}_{l}(r) u_{l+1}(r) r dr \right) +  \\ 
& \left. \left. 
B_{l,m}^{^\ast \alpha}({\bf {k+G^{\prime}}}) 
B_{l+1,m+1}^{\alpha}({\bf {k+G}})
\left( \int\limits_0^{R_{\alpha}} \dot{u}_{l}(r) \dot{u}^{\prime}_{l+1}(r) r^2 dr +
 (l+2) \int\limits_0^{R_{\alpha}} \dot{u}_{l}(r) \dot{u}_{l+1}(r) r dr \right) 
\right \}
\right \} 
\end{array}
\nonumber\\
\end{eqnarray}
To obtain the atomic-sphere contributions to the momentum matrix elements, Eqs. (\ref{Phi_x+y}-\ref{Phi_z}) have to be multiplied with the variational
coefficients and summed up over all basis functions:
\begin{eqnarray}
\langle n^{\prime}{\bf k} \left| \nabla_x \right| n{\bf k} \rangle_{\mathrm{MT}_{\alpha}}
& = &
\frac{1}{2 \phantom{i}} 
\sum_{{\bf G}^{\prime},{\bf G}}
C_{n^{\prime} {\bf k}}^{\ast} ({\bf G^{\prime}})
\left(
^{\alpha}\Phi_{{\bf {k+G^{\prime}},\bf {k+G}}}^{x+iy} +
^{\alpha}\Phi_{{\bf {k+G^{\prime}},\bf {k+G}}}^{x-iy} \right)
C_{n {\bf k}} ({\bf G}) \nonumber \\
\langle n^{\prime}{\bf k} \left| \nabla_y \right| n{\bf k} \rangle_{\mathrm{MT}_{\alpha}} & = &
\frac{1}{2i} 
\sum_{{\bf G}^{\prime},{\bf G}}
C_{n^{\prime} {\bf k}}^{\ast} ({\bf G^{\prime}})
\left(
^{\alpha}\Phi_{{\bf {k+G^{\prime}},\bf {k+G}}}^{x+iy} -
^{\alpha}\Phi_{{\bf {k+G^{\prime}},\bf {k+G}}}^{x-iy} \right)
C_{n {\bf k}} ({\bf G}) \nonumber \\
\langle n^{\prime}{\bf k} \left| \nabla_z \right| n{\bf k} \rangle_{\mathrm{MT}_{\alpha}} & = &
\phantom{\frac{1}{2i}}
\sum_{{\bf G}^{\prime},{\bf G}}
C_{n^{\prime} {\bf k}}^{\ast} ({\bf G^{\prime}})
^{\alpha}\Phi_{{\bf {k+G^{\prime}},\bf {k+G}}}^{z}
C_{n {\bf k}} ({\bf G}) 
\label{ME_alpha}
\end{eqnarray}
\subsubsection{Contributions from the interstitial region}
With the planewave basis (\ref{basis_Int}) the interstitial contribution to the matrix elements (\ref{ME_sum}) can be easily worked out:
\begin{equation}
\langle n^{\prime}{\bf k} \left| \nabla \right| n{\bf k} \rangle_{\mathrm{I}} =
\frac{1}{\Omega_c} \sum_{{\bf G}^{\prime},{\bf G}}
C_{n^{\prime} {\bf k}}^{\ast} ({\bf G^{\prime}})
C_{n {\bf k}} ({\bf G}) 
\int\limits_{\mathrm{I}}
e^ {i {\bf { \left ( G^{\prime} - G \right ) }{\bf r} } } d{\bf r} 
\label{ME_Int}
\end{equation}
The integration in Eq. (\ref{ME_Int}) over the interstitial region is carried out by integrating over the whole unit cell and subtracting the integral over the atomic spheres (see Appendix). This procedure finally leads to
\begin{eqnarray}
\langle n^{\prime}{\bf k} \left| \nabla \right| n{\bf k} \rangle_{\mathrm{I}} =
\frac{i}{\Omega_c} \sum_{{\bf G}} 
\left ( {\bf k+G} \right) C_{n {\bf k}} ({\bf G})
\left[
C_{n^{\prime} {\bf k}}^{\ast} ({\bf G}) \left( \Omega_c - \sum_{\alpha} V_{\alpha} \right) -
\sum_{\alpha} 3 V_{\alpha} 
\nonumber \right.\\
\left.\sum_{{\bf G}^{\prime} \neq {\bf G}} 
C_{n^{\prime} {\bf k}}^{\ast} ({\bf G}^{\prime}) 
\frac{\sin \left( \left| {\bf G}^{\prime}-{\bf G} \right| R_{\alpha} \right)
   - (\left| {\bf G}^{\prime}-{\bf G} \right| R_{\alpha} )
    \cos \left( \left| {\bf G}^{\prime}-{\bf G} \right| R_{\alpha} \right)}
    {\left (\left | {\bf G}^{\prime} - {\bf G} \right | R_{\alpha} \right)^3} 
 e^{i \left( {\bf G} - {\bf G}^{\prime} \right) {\bf S_{\alpha}}} \right]\;,
\label{ME_Interstitial}
\end{eqnarray}
where $V_{\alpha}$ is the volume of the atomic sphere $\alpha$.
\section{Results}

\subsection{Aluminum}
\begin{figure}[htb]
\includegraphics[width=12cm]{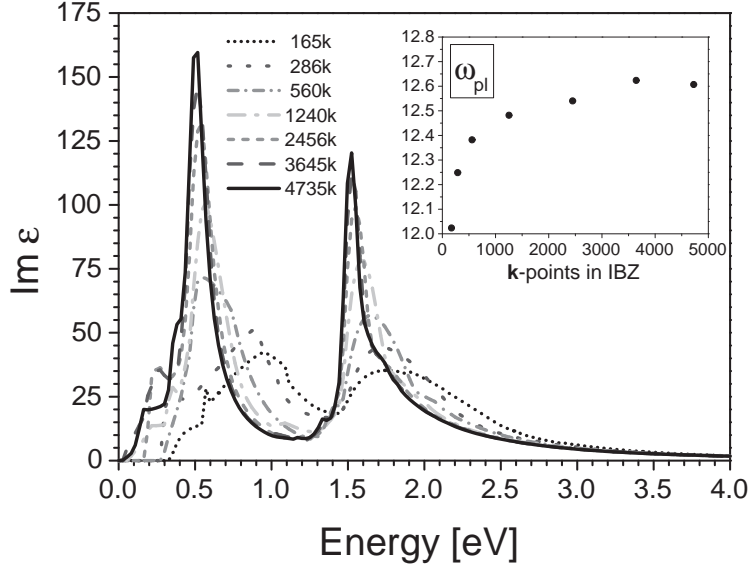}
\caption{
  Imaginary part of the frequency dependent dielectric function of Al
  for a series of {\bf k} point meshes. In the inset the dependence of
  the plasma frequency (in eV) on the {\bf k} point sampling is
  depicted.}
\label{fig:Imeps_mesh}
\end{figure}
As the first example fcc aluminum has been chosen. It has been
intensively studied in literature,
\cite{Ehrenreich:63,Shiles:80,Szmulowicz:81,Smith:86,Lee:94} where 
the most puzzling experimental problem at that time was the distinction
between interband and intraband contributions to the absorption process.
Here, it serves as a test case mainly in terms of convergence with 
respect to the most important convergence parameters. While the results 
did not turn out to be sensitive to the number
of LAPW's, there is a crucial dependence on the BZ sampling as already 
found by Lee and Chang.\cite{Lee:94} 
Fig.~\ref{fig:Imeps_mesh} shows the interband contribution to the
imaginary part of the dielectric function (Eq. \ref{eq:eintercomp}) for 
a series of ${\bf k}$ point meshes. The peaks do not only get sharper 
with increasing number of ${\bf k}$ points, but they also exhibit a
pronounced energy shift. Only with the most dense mesh quantitative
agreement with experimental data is achieved.\cite{remark} The inset 
of Fig. \ref{fig:Imeps_mesh} exhibits the plasma frequency  
(Eq. \ref{eq:omegapl}) as a function of the number of ${\bf k}$
points in the irreducible part of the Brillouin zone (IBZ). Also in
this case highest accuracy is needed to reach the converged value of
$\omega_{pl}=12.6\:eV$.  This sensitivity can be understood by the
high symmetry of the crystal structure and the simplicity of the
material. Since the main contributions stem from certain regions of
the BZ,\cite{Szmulowicz:81} a refinement of the mesh there can
dramatically change this contribution and -- due to the high weight in
the BZ -- the total spectrum. In contrast, in more complex materials
the optical absorption usually represents a sum of many different
contributions from several regions in the BZ and many band
combinations. Therefore a change of one or some of these
contributions cannot affect the total spectrum that much.
\begin{figure}[htb]
\includegraphics[width=12cm]{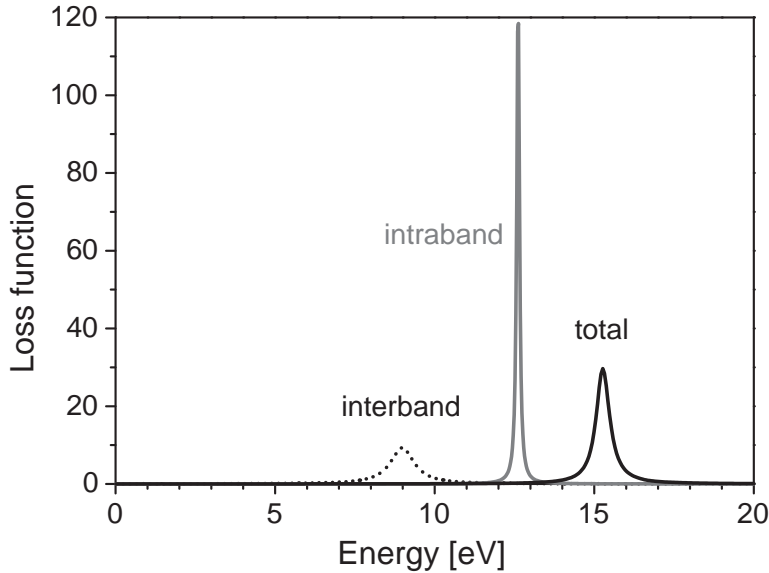} 
\caption{
  Loss function for Al (full line) calculated with 4735 {\bf k} points
  in the IBZ.  The dashed (dotted) line shows the loss function of the
  pure intraband (interband) contribution.}
\label{fig:Al_eloss}
\end{figure}
Fig. \ref{fig:Al_eloss} illustrates the separation of interband and
intraband contributions to the optical spectra. For this purpose the
total loss function is plotted (full line), but also the corresponding
functions if there was interband (dotted line) or intraband
(dashed line) contributions only. Judging from the shape of the curves
all of them could be interpreted as free electron behavior but with
very different plasma frequencies. Therefore the plasma frequency was
for a long time thought to be 15.2\:eV which is the peak position of
the total spectrum. For a more detailed discussion of the problem see
Ref.  \onlinecite{Smith:86}.
\begin{figure}[htb]
\includegraphics[width=12cm]{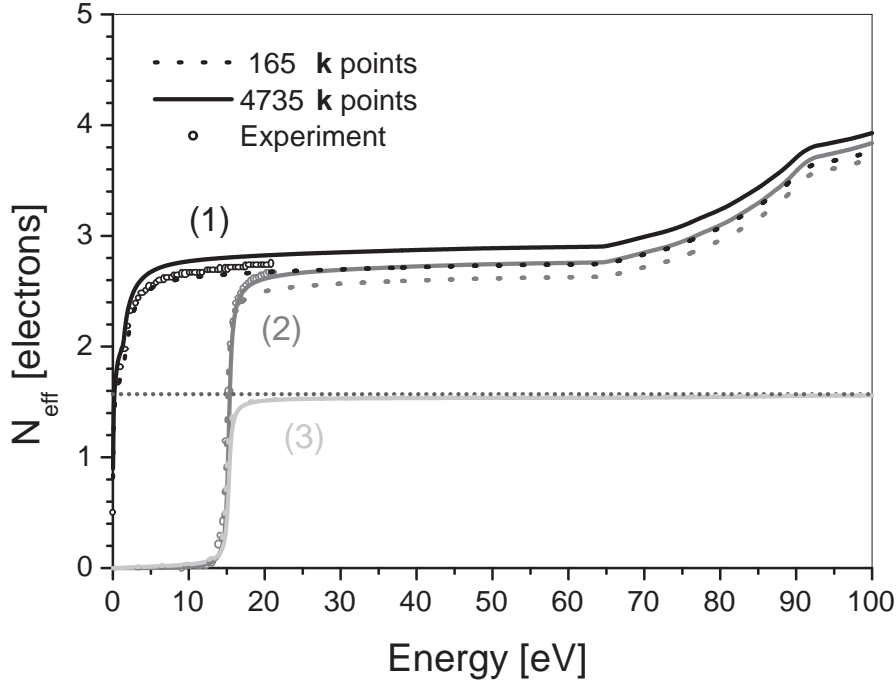} 
\caption{Sumrule check for the optical spectra of Al by carrying out the integrations given in Eqs. \ref{sumrule1} (1), \ref{sumrule2} (2), and \ref{sumrule3} (3) as a function of the upper integration limit (energy $\hbar\omega$ in eV). The calculated values are given for two different numbers of ${\bf k}$ points in the BZ, i.e. 165 (dashed lines) and 4735 (full lines). The corresponding experimental data (open circles) are taken from Ref. \onlinecite{Ehrenreich:63}. The dotted line indicates $\pi$/2 for comparison with the third sumrule (3).}
\label{fig:Al_sumrules}
\end{figure}
Finally the sumrules provide another test for the quality of the
calculations.  To this extent, all three sumrules as given in Eqs.
(\ref{sumrule1} -- \ref{sumrule3}) have been tested. In Fig.
\ref{fig:Al_sumrules} the results of the corresponding integrations
are depicted as a function of energy, i.e. the upper integration
limit. The theoretical results for sumrules \ref{sumrule1} and
\ref{sumrule2} excellently reproduce the region where experimental
data are available, while the third curve approaches $\pi$/2 within
0.1 \%. Having a closer look to sumrule (1), the kink in the low
energy range indicates the setting in of the interband contributions.

\subsection{Gold}
%
%

The electronic bands have been studied very intensively in the
literature,\cite{Christensen:71,Christensen:76,Takeda:80,Bross:00} and
much attention has been paid to relativistic effects on the band
structure and henceforth the optical properties.\cite{Christensen:71}
But the optical spectra have not been investigated in detail from the 
first-principles side. Therefore they shall be studied in more detail
 here. 

A lattice parameter of 7.66 a.u. has been taken which is the optimized
value with respect to a scalar-relativistic LDA calculation. The
self-consistency cycles have been performed with 
65 ${\bf k}$ points in the IBZ. The BZ integrations for obtaining the 
density of states (DOS), joint DOS, and the optical properties used in 
the figures were carried out with 1240 points in the IBZ. (For 
convergence tests see below.)   
In order to investigate relativistic effects we have performed {\it 
non-relativistic}, {\it scalar-relativistic}, and {\it relativistic} 
calculations, where in the latter spin-orbit coupling has been treated 
by a second variational scheme. In all cases the core states are 
calculated fully relativistically by solving the Dirac equation. We 
treat the semi-core states 5$s$, 5$p$ and 4$f$ in the valence region by 
local orbitals. Therefore these low lying states cannot be expected to 
be very well described in the non-relativistic case. Our procedure, 
however, has the advantage that the optical properties can be handled by
the formalism described above up to high excitation energies which is 
needed for the Kramers--Kronig analysis as well as for the test of the 
sumrules. At the same time the influence of relativistic effects can be 
studied in detail. 

Also in the case of gold, we have paid attention to the convergence of 
the optical spectra with respect to the ${\bf k}$ point sampling. It 
turned out that the density of the ${\bf k}$ point mesh is much less 
important than in aluminum. Already rather course meshes give all 
results with high accuracy. For example, the plasma frequency of the 
relativistic calculation changes by 0.1\:eV only, i.e. from 8.966 to 
8.867, when increasing the number of ${\bf k}$ points in the IBZ from 
165 to 11480. In fact, already with 47 ${\bf k}$ points a value of 
8.866\:eV is obtained. 
For the interband contributions, the spectra above 2\:eV can hardly be 
distinguished when plotted for different BZ samplings. Only the first 
interband peak (below 2\:eV) in the real part of the dielectric function
is somewhat broader with more spectral weight at lower energies when 
the number of irreducible ${\bf k}$ points is reduced below 1240. 
\begin{figure}[htb]
\includegraphics[width=12cm]{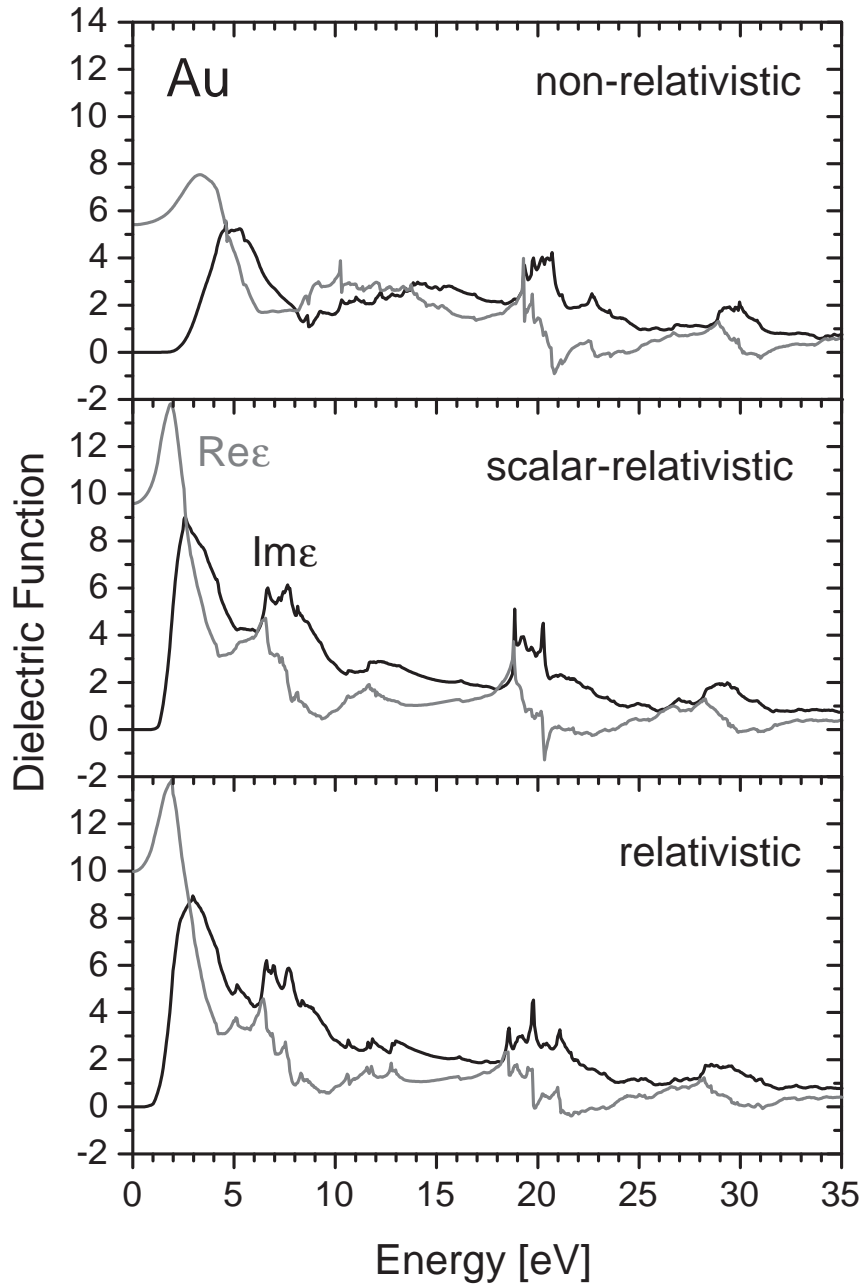} 
\caption{Interband contribution to the dielectric function of gold: real (grey lines) and imaginary part (black lines) of $\mathrm{Im}\:\epsilon$ from non-relativistic (top), scalar-relativistic (middle) and relativistic calculations (bottom).}
\label{fig:Au_eps}
\end{figure}
\begin{figure}[htb]
\includegraphics[width=18cm]{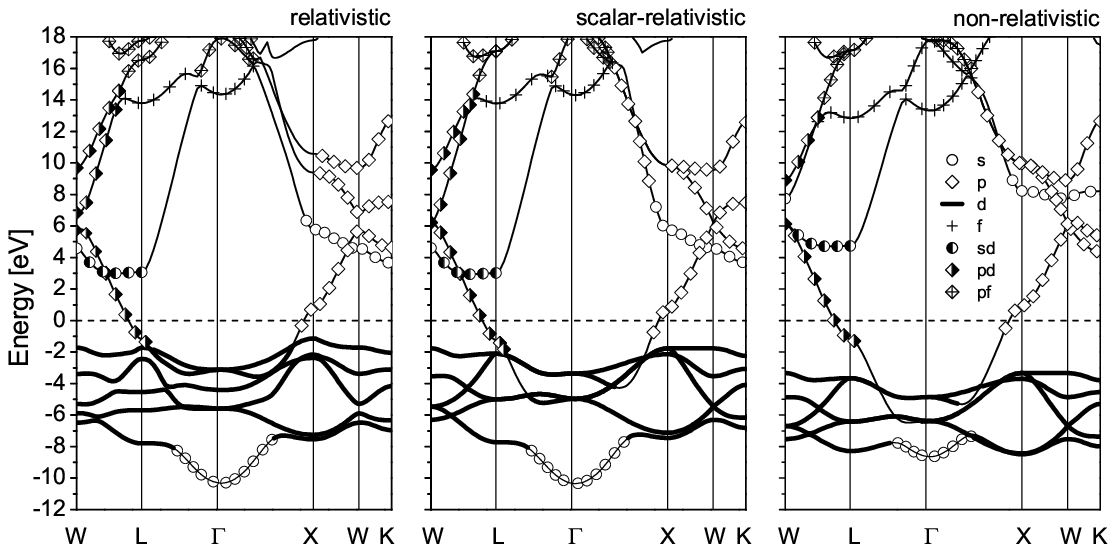} 
\caption{Relativistic effects in the band structure of gold: 
non-relativistic (right), scalar-relativistic (middle) and relativistic 
calculations (left) have been performed. The main $l$-like band 
character is highlighted by symbols as indicated in the legend. The thin
lines without symbols represent states with more than two different significant contributions.}
\label{fig:Au_bands}
\end{figure}
In Fig. \ref{fig:Au_eps} the imaginary part of the dielectric function arising from interband transitions is shown for the three cases mentioned above. While there is hardly any difference between the scalar-relativistic calculation and the treatment including spin-orbit coupling the biggest change is observed when taking into account the scalar-relativistic effects which are due to the Darwin shift and the mass--velocity term.
 
\begin{figure}[htb]
\begin{center}
\includegraphics[width=12cm]{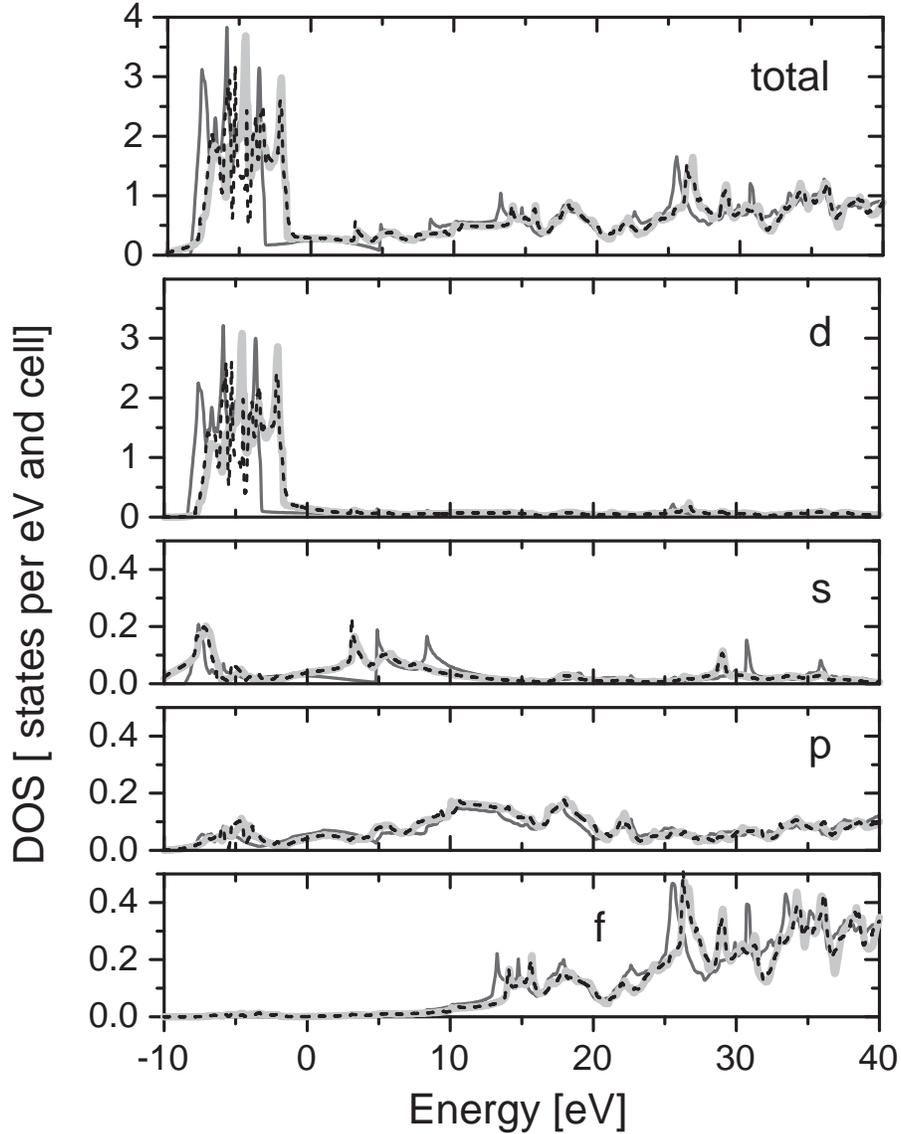} 
\caption{Total and site-dependent densities of states of Au from a non-relativistic (dark grey), a scalar-relativistic (light grey) and a  relativistic calculation (black).
spin-orbit coupling}
\label{fig:Au_dos}
\end{center}
\end{figure}

In order to analyze the origin of theses effects the corresponding band structures are shown in Fig. \ref{fig:Au_bands}. One striking scalar-relativistic effect is the broadening and upward shift of the filled 5$d$ bands, which is much bigger than the splitting of some states due to spin-orbit coupling. Also the 5$s$ states lying at -9\:eV at the $\Gamma$ point in the non-relativistic case exhibit an increased band width by more than a factor of two. Relativistic effects of the 6$s$ and 5$p$ states in the conduction band are highlighted in the figure by indicating the predominant band character along the high symmetry lines X--W and W--K. The 5$f$ states dominate the band structure above 12\:eV in the non-relativistic case, and these bands are pushed up by roughly 1.5\:eV. All these effects can be seen more clearly in the total and site-projected densities of states depicted in Fig. \ref{fig:Au_dos}. 

The first half of the spectrum is dominated by transitions between $d$ 
and $p$ like states which generally are shifted to lower energies due to
the upward shift of the $d$ band when relativistic terms are included in
the calculations.
In particular, the non-relativistic $\mathrm{Im}\:\epsilon$ centered 
around 5\:eV is created by 5$d$ to 6$p$ transitions with the partially 
filled band of mainly $p$ character providing the final states. This 
peak is shifted down by 2\:eV in the scalar-relativistic case due to the
nearly rigid upward shift of the 5$d$ bands. The increase of the 
spectral weight can be traced back to the $1/\omega^2$ behavior of 
$\mathrm{Im}\:\epsilon$. 
The double peak structure around 6-8\:eV visible in the relativistic 
cases come from transitions from the three highest lying valence levels 
to the first two unoccupied bands. They also exhibit mainly $d$ to $p$ 
character, but due to some $p$ admixture in the high lying valence 
states and a considerable amount of $d$ character in the final states 
also finite $p$ to $d$ contributions are present. 
Also the $d$ to $f$ transitions around 20\:eV are shifted to lower 
energies because the $f$ states move up only half as much as the $d$ 
states. 
Including spin orbit coupling only very minor changes are obtained. 
In particular, the $d$ to $f$ transitions are broadened reflecting a splitting of the $d$ bands as can be seen in the band structure and the symmetry projected DOS. The transitions in this case exhibit an asymmetry with respect to the Fermi level. In all cases the initial states are mainly the $d$ bands between -10 and -1\:eV, while the final states range from the Fermi level up to very high energies.
\begin{figure}[htb]
\includegraphics[width=12cm]{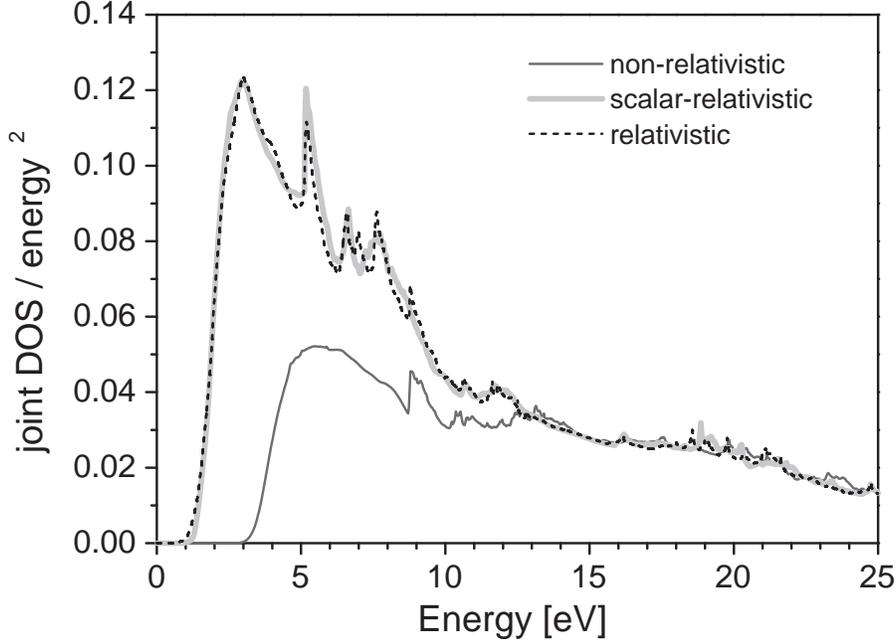} 
\caption{Joint density of states of gold divided by energy squared. The data are shown for the non-relativistic case (dark gray full line), the scalar-relativistic case (light grey full lines) and the relativistic case (black scattered line).}
\label{fig:Au_jdos}
\end{figure}
As already pointed out by Christensen\cite{Christensen:71} the matrix element effect is relatively small. The imaginary part of the dielectric function is already very well estimated by the joint density of states as can be seen in Fig. \ref{fig:Au_jdos}. Since the dielectric function suppresses higher energy transitions due to its $1/\omega^2$ behavior (Eq. \ref{eq:dieltens}) we have divided the joint DOS by the energy squared. Only a few features there are canceled by selection rules, like the sharp peak at 4.5\:eV arising from the parallel bands of $d$ character at -2\:eV and $s$ character at 2.5\:eV as clearly seen along the line W--L in the BZ. In the non-relativistic case up to 13\:eV, the magnitude is much too small and the peaks are found at too high energies as already seen in the optical spectra.
\begin{figure}[htb]
\includegraphics[width=12cm]{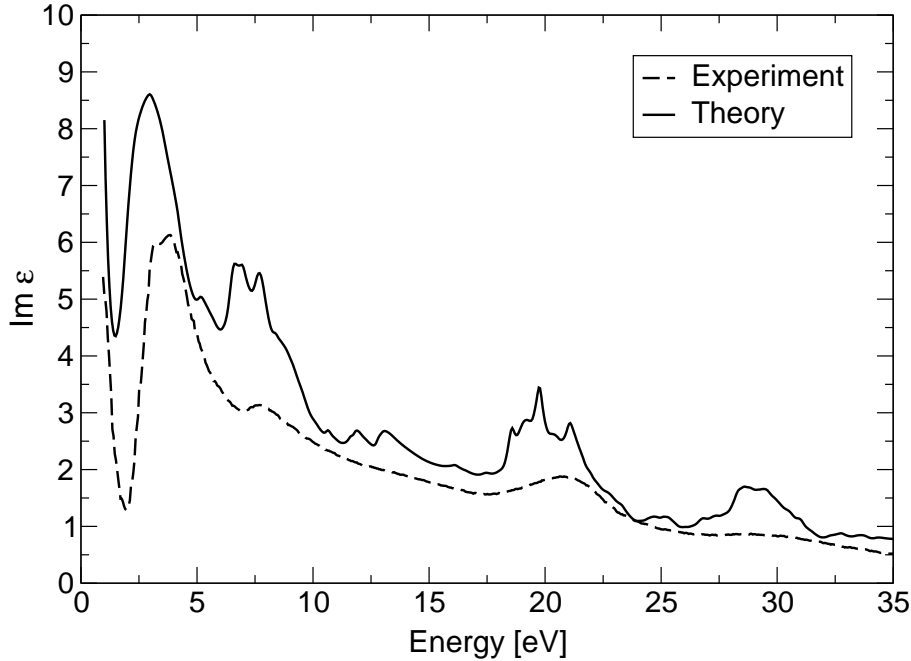} 
\caption{Imaginary part of the dielectric function for gold including a life-time broadening of 0.1eV obtained by a relativistic calculation compared to the corresponding experiment taken from Cooper et al.\cite{Cooper:65}}
\label{fig:Au_Im_eps}
\end{figure}
\begin{figure}[htb]
\includegraphics[width=12cm]{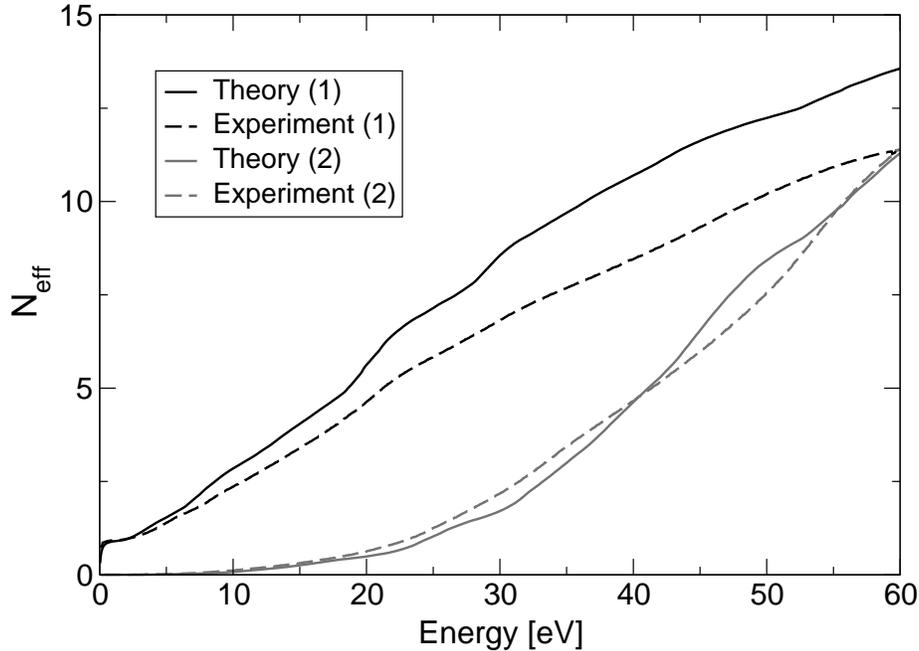} 
\caption{Sumrule test for the optical spectra of gold obtained by a relativistic calculation compared to the corresponding experiment taken from Cooper et al.\cite{Cooper:65} The numbers in brackets refer to the corresponding sumrules (Eqs. (\ref{sumrule1} and \ref{sumrule2}).}
\label{fig:Au_sumrules}
\end{figure}
Comparing our results to experimental data (Fig. \ref{fig:Au_Im_eps}),
the overall agreement is good. There is two main points which have to
be considered. First, the experimental spectrum is in general smaller
in magnitude. This seems to be a common problem which can most
probably be traced back to a diminished reflectivity at the sample
surface.\cite{Kathrin:04} Second, although all experimental features
are reproduced by theory, the peak positions appear to be slightly too
low in the latter. Since most of the transitions contain $d$ states
either as initial or as final state, the width of the $d$ band is 
presumably the source of this shortcoming. The number of effective
charges $N_{eff}$ probed by the two sumrules (Eqs.(\ref{sumrule1}) and
(\ref{sumrule2}) behave quite differently as can be seen in Fig.
\ref{fig:Au_sumrules}. While the integration of ${\mathrm Im}\epsilon$
reflects the different magnitudes of theoretical and experimental
data, the integration over the loss function is much less sensitive
showing very good agreement between theory and experiment in the whole
energy range.

Before the over-interpretation of KS states are blamed for
the small discrepancies, other reasons should be investigated.
From the theoretical side they can be found in the correlation effects
being underestimated by the LDA as discussed above, but also in the
rather poor basis set of linearized methods for the conduction bands.
Furthermore, for the high energy photons the $q=0$ limit is not
justified any more and thus a finite momentum transfer should be
considered in future work. But also the experimental side gives rise 
to shortcomings, as e.g. the Kramers--Kronig analysis requires a
continuation of the reflectance spectrum. We need to include
$\mathrm{Im}\:\epsilon$ up to more than 60\:eV in the KK
transformation in order to obtain $\mathrm{Re}\:\epsilon$ and
henceforth the loss function up to 30\:eV reliably.  

Although the
changes of the band crossing the Fermi level are hardly visible from
the band structure, the plasma frequency changes from 9.02 to 9.51
when taking into account scalar-relativistic effects and becomes 8.86
when spin-orbit coupling is introduced. This sensitivity of this
quantity is a measure for the reliability of the Fermi surface by the
fully-relativistic calculation. $\omega_{pl}$ excellently agrees with
its experimental counterpart of 8.83.\cite{Cooper:65} An explanation
can also be given for the fact that the plasma frequency hardly
depends on the {\bf k} point mesh. Since the only band contributing to
the effective number of free carriers has a linear shape around the
crossing with the Fermi level the linear tetrahedron method used in our
work turns out to be a very effective tool for the Fermi-surface
integration.

\section{Conclusions}
We have provided a scheme for the calculation of linear optical
 properties of solids in the random-phase approximation. It is worked 
out within the LAPW method which does not require approximations in the 
solution of the KS equation and thus is a good starting point for the
evaluation of the KS states in terms of band structure and excited
states.

Aluminum, which is a prototype for a free-electron-like metal, exhibits
excellent agreement with experiments in terms of plasma frequency, peak
positions of the interband transitions, as well as sumrules. This
agreement can, however, only be achieved when the BZ integration is
carried out with extreme accuracy. Thus highest precision is in general
needed to draw conclusions about the quality of the KS states.

For gold, the optical properties are less sensitive to the BZ sampling.
The plasma frequency is in excellent agreement with experiment when
relativistic effects are taken into account including spin-orbit
coupling. The latter is, however, less important for the interband
transitions than the scalar-relativistic contributions, i.e. the Darwin
shift and the mass--velocity term. 
Shortcomings can at least be partly traced back to the approximations 
used for the ground state calculation, i.e. the LDA which does not 
properly account for correlation effects and thus cannot bring out the
$d$ band width and location well enough. Other possible weak points are
the insufficient basis set for the unoccupied states, the $q=0$ limit 
for high energy transfer, as well as the missing local-field effects. 
The latter, however, turned out to be of minor importance for the 
calculation of hot-electron lifetimes.\cite{Ladstaedter:03} 
All these facts have to be accounted for before a final statement about 
the interpretability of KS states for a certain material can be made. 
Nevertheless, in both cases studied here, they provide at least a very 
good first approximation to single state and excited state energies, 
respectively.

\bigskip
\noindent
{\bf Acknowledgments}

We appreciate support from the Austrian Science Fund (FWF), project
P16227-PHY, and by the EU research and training network
{\it EXCITING}, contract number HPRN-CT-2002-00317. The work was supported in part by the Materials Simulation Center, a Penn-State MRSEC and MRI facility.

\section{Appendix}

\subsection{k.p Perturbation Theory}
\label{sec:kdp}

In the position representation, the one-electron wavefunction
$\langle \mathbf{r} \left| \right. n,\mathbf{k}\rangle = \psi _{n,{\bf
    k}}({\bf r})$ is the solution of the Schr\"{o}dinger equation
\begin{equation}
\left[ {p^{2}\over 2m} + V({\mathbf r}) \right] 
\psi_{n,{\bf k}}({\bf r}) = 
\varepsilon_{n,{\bf k}} \psi_{n,{\bf k}}({\bf r})\;.
\label{KS-equation_2}
\end{equation}
Since the crystal potential $V(\mathbf{r})$ has the periodicity of the
lattice, using Bloch's theorem we can write
\begin{equation}
  \label{eq:bloch}
  \psi_{n,{\mathbf k}}({\mathbf r}) = e^{i \mathbf k\cdot \mathbf
  r}\: u_{n,{\mathbf k}}({\mathbf r})\;.
\end{equation}
The periodic part of the wavefunction 
$u_{n,\mathbf{k}}(\mathbf{r})$ obeys
\begin{equation}
  \label{eq:ueq}
  H_{\mathbf{k}}\; u_{n,\mathbf{k}}(\mathbf{r})=\varepsilon_{n,{\bf
  k}}\; u_{n,\mathbf{k}}(\mathbf{r})\; ,
\end{equation}
where 
\begin{equation}
  \label{eq:hk}
H_{\mathbf{k}}= {\left(\mathbf{p}+\mathbf{k}\right)^2\over 2m} +
V(\mathbf{r})\;. 
\end{equation}
We want to obtain an expression for 
$u_{n,\mathbf{k}+\mathbf{q}}(\mathbf{r})$ and 
$\varepsilon_{n,\mathbf{k}+\mathbf{q}}$ valid for small $\mathbf{q}$
in terms of the values for $\mathbf{q}=0$ by perturbation theory.
The equation for $u_{n,\mathbf{k}+\mathbf{q}}(\mathbf{r})$ is
\begin{equation}
  \label{eq:ukqeq}
  H_{\mathbf{k}+\mathbf{q}}\; u_{n,\mathbf{k}+\mathbf{q}}(\mathbf{r})=
\varepsilon_{n,{\bf k}+\mathbf{q}}\; 
u_{n,\mathbf{k}+\mathbf{q}}(\mathbf{r})\; ,
\end{equation}
with 
\begin{equation}
  \label{eq:hkq}
H_{\mathbf{k}+\mathbf{q}}= H_{\mathbf{k}} + {\hbar\mathbf{p}\cdot\mathbf{q}\over m}+
{\hbar^2\mathbf{k}\cdot\mathbf{q}\over m}+
{\hbar^2 q^2\over 2 m}\;, 
\end{equation}
where the last three terms can be treated as a perturbation.

At this point it will be useful to introduce the notation 
\begin{equation}
  \label{eq:ur}
  \left(\mathbf{r}\left|\right.n,\mathbf{k}\right)=
  u_{n,\mathbf{k}}(\mathbf{r})
\end{equation}
and 
\begin{equation}
  \label{eq:blochme}
  \left(n^\prime ,\mathbf{k}^\prime \left| \hat{O} \right|
  n,\mathbf{k}\right) = {1\over\Omega}\int_{\Omega} d\mathbf{r}\:
  u^\ast_{n^\prime,\mathbf{k}^\prime}(\mathbf{r})\:
  \hat{O}\:u_{n,\mathbf{k}}(\mathbf{r})\;,
\end{equation}
for any operator $\cal{O}$ with the integral running over the unit
cell of volume $\Omega$.

According to this notation, the wavefunction to first order in
$\mathbf{q}$ for a non-degenerate state is
\begin{equation}
  \label{eq:wfq}
  \left|n,\mathbf{k}+\mathbf{q}\right) =
  \left|n,\mathbf{k}\right) +
  \sum_{n^{\prime}\neq n} 
  \left|n^{\prime},\mathbf{k}\right)
  { \left(n^{\prime},\mathbf{k}\left| 
    H_{\mathbf{k}+\mathbf{q}}-H_{\mathbf{k}}
    \right| n,\mathbf{k}\right) \over
    \varepsilon_{n,\mathbf{k}} -
    \varepsilon_{n^{\prime},\mathbf{k}} 
  }
\end{equation}
Only one term from the perturbation Hamiltonian gives a contribution
different from zero to the first order correction of the wavefunction
\begin{equation}
  \label{eq:wfqq}
  \left|n,\mathbf{k}+\mathbf{q}\right) =
  \left|n,\mathbf{k}\right) +
  \sum_{n^{\prime}\neq n} 
  \left|n^{\prime},\mathbf{k}\right)
  { \left(n^{\prime},\mathbf{k}\left| 
    \frac{\hbar}{m}\mathbf{p}\cdot\mathbf{q}
    \right| n,\mathbf{k}\right) \over
    \varepsilon_{n,\mathbf{k}} -
    \varepsilon_{n^{\prime},\mathbf{k}} 
  }
\end{equation}
To linear order in $\mathbf{q}$ the expression for the energy is
\begin{equation}
  \label{eq:enq}
  \varepsilon_{n,\mathbf{k}+\mathbf{q}} =
  \varepsilon_{n,\mathbf{k}} +
\left(n,\mathbf{k}\left| 
    \frac{\hbar\:\mathbf{p}\cdot\mathbf{q}}{m}+
    \frac{\hbar^2\:\mathbf{k}\cdot\mathbf{q}}{m}
    \right| n,\mathbf{k}\right)\;.
\end{equation}
Since the states are normalized this expression can be written as
\begin{equation}
  \label{eq:enqq}
  \varepsilon_{n,\mathbf{k}+\mathbf{q}} =
  \varepsilon_{n,\mathbf{k}} +
\frac{\hbar}{m} \left[
\left(n,\mathbf{k}\left| 
    \mathbf{p}\right| n,\mathbf{k}\right) + \hbar\:\mathbf{k}
 \right]\cdot\mathbf{q}\;
\end{equation}

The momentum matrix element evaluated in Section \ref{sec:mme} can be
expressed as
\begin{equation}
  \label{eq:mme}
  \mathbf{p}_{l,n,\mathbf{k}} \equiv
\langle l,\mathbf{k}\left| \mathbf{p} \right| n,\mathbf{k}
\rangle = 
\delta_{l,n}\;\hbar\:\mathbf{k} + 
\left( l,\mathbf{k}\left| \mathbf{p} \right| n,\mathbf{k}\right)\;.
\end{equation}
In terms of this definition, the wavefunctions and energies to first
order in $\mathbf{q}$ are given by 
\begin{equation}
  \label{eq:wfqp}
  \left|n,\mathbf{k}+\mathbf{q}\right) =
  \left|n,\mathbf{k}\right) + \frac{\hbar}{m}
  \sum_{n^{\prime}\neq n} 
  \left|n^{\prime},\mathbf{k}\right)
  {  
    \mathbf{p}_{n^{\prime},n,\mathbf{k}}
    \over
    \varepsilon_{n,\mathbf{k}} -
    \varepsilon_{n^{\prime},\mathbf{k}} 
  }  \cdot\mathbf{q}
\end{equation}
and
\begin{equation}
  \label{eq:enqp}
  \varepsilon_{n,\mathbf{k}+\mathbf{q}} =
  \varepsilon_{n,\mathbf{k}} +
\frac{\hbar}{m}\: \mathbf{p}_{n,n,\mathbf{k}}\:
\cdot\mathbf{q}\;  
\end{equation}

\subsection{Matrix elements for small $q$}
\label{sec:mmee}

In this appendix we use the $\mathbf{k}\cdot\mathbf{p}$
expressions developed in Appendix \ref{sec:kdp} to evaluate the matrix
elements for $\mathbf{q}\to 0$. We first write the matrix elements in
terms of the periodic part of the wavefunction
\begin{equation}
  \label{eq:meu}
  M^0_{l,n}(\mathbf{k},\mathbf{q}) =
\langle l,\mathbf{k}\left| 
e^{-i\:\mathbf{q}\cdot\mathbf{r}}
\right|n,\mathbf{k}+\mathbf{q}\rangle =
\left( l,\mathbf{k}\left|
\right.n,\mathbf{k}+\mathbf{q}\right)\;.
\end{equation}
Using the expression of the wavefunction for small $\mathbf{q}$ of
Eq.(\ref{eq:wfqp}), the matrix element at lowest order is
\begin{equation}
  \label{eq:meqp}
    M^0_{l,n}(\mathbf{k},\mathbf{q}\to 0) =
\delta_{l,n}+(1-\delta_{l,n})
\frac{\hbar}{m}\:
  {  
    \mathbf{p}_{l,n,\mathbf{k}}
    \over
    \varepsilon_{n,\mathbf{k}} -
    \varepsilon_{l,\mathbf{k}} 
  }  \cdot\mathbf{q}
\end{equation}

\subsection{Relations between spherical harmonics \label{A_harmonics}}
Below, several well known relations between the spherical harmonics are given, which are needed to evaluate the momentum matrix elements:
\begin{eqnarray}
e^{+ i\varphi} \sin \theta Y_{l,m} & 
             = &  F_{l,m}^{(1)} Y_{l+1,m+1} +  F_{l,m}^{(2)} Y_{l-1,m+1}\\
e^{- i\varphi} \sin \theta Y_{l,m} & 
             = &  F_{l,m}^{(3)} Y_{l+1,m-1} +  F_{l,m}^{(4)} Y_{l-1,m-1}\\
\cos \theta Y_{l,m} & = &  F_{l,m}^{(5)} Y_{l+1,m} + F_{l,m}^{(6)} Y_{l-1,m}
\label{A_Y_lm}
\end{eqnarray}
with
\begin{eqnarray}
F_{l,m}^{(1)} & = - &  \sqrt{ \frac{(l+m+1)(l+m+2)}{(2l+1)(2l+3)} }\\
F_{l,m}^{(2)} & =   &  \sqrt{ \frac{(l-m)(l-m-1)}{(2l-1)(2l+1)}   }\\
F_{l,m}^{(3)} & =   &  \sqrt{ \frac{(l-m+1)(l-m+2)}{(2l+1)(2l+3)} }\\
F_{l,m}^{(4)} & = - &  \sqrt{ \frac{(l+m)(l+m-1)}{(2l-1)(2l+1)}   }\\
F_{l,m}^{(5)} & =   &  \sqrt{ \frac{(l-m+1)(l+m+1)}{(2l+1)(2l+3)} }\\
F_{l,m}^{(6)} & =   &  \sqrt{ \frac{(l-m)(l+m)}{(2l-1)(2l+1)}     }
\label{A_F_lm}
\end{eqnarray}
\begin{eqnarray}
e^{+ i\varphi}  \left ( \cos \theta \frac {\partial} {\partial \theta} +
\frac{i}{\sin \theta} \frac {\partial} {\partial \varphi} \right ) Y_{l,m}
         & = &  -l F_{l,m}^{(1)} Y_{l+1,m+1} +  (l+1) F_{l,m}^{(2)} Y_{l-1,m+1}\\
e^{- i\varphi}  \left ( \cos \theta \frac {\partial} {\partial \theta} -
 \frac{i}{\sin \theta} \frac {\partial} {\partial \varphi} \right ) Y_{l,m}
         & = &  -l F_{l,m}^{(3)} Y_{l+1,m-1} +  (l+1) F_{l,m}^{(4)} Y_{l-1,m-1}\\
- \sin \theta \frac {\partial} {\partial \theta} Y_{l,m}  
         & = & -l F_{l,m}^{(5)} Y_{l+1,m+1} +  (l+1) F_{l,m}^{(6)} Y_{l-1,m+1}
\label{A_dY_lm}
\end{eqnarray}
\subsection{The step function}
The analytic evaluation of a planewave integral over the interstitial 
region as needed in Eq. (\ref{ME_Int}) is carried out by integrating over the whole unit cell and subtracting the contributions of the atomic spheres, which are
\begin{equation}
\int\limits_{MT_{\alpha}}
e^ {i {\bf { \left ( {\bf G}^{\prime} - {\bf G} \right ) }
{\bf r} } } d{\bf r} =
\left\{
\begin{array}{ll}
V_{\alpha} 
& \quad \quad  {\bf G}^{\prime} = {\bf G}  \\
3 V_{\alpha} 
\frac{\sin \left( R_{\alpha} \left| {\bf G}^{\prime}-{\bf G} \right| \right)
    - \cos \left( R_{\alpha} \left| {\bf G}^{\prime}-{\bf G} \right| \right)}
    {\left( R_{\alpha}  \left| {\bf G}^{\prime} - {\bf G} \right | \right)^3} 
e^ {i \left( {\bf G}^{\prime} - {\bf G} \right) {\bf S_{\alpha}} }
& \quad \quad   {\bf G}^{\prime} \neq {\bf G}
\end{array}
\right. 
\end{equation}
\medskip
It utilizes the Rayleigh-expansion of a planewave in terms of
spherical harmonics:
\begin{equation}
e^{i{\bf {Gr}} } = 4\pi e^{i{\bf {GS} }_{\alpha}} \sum_{lm} i^{l} j_{l}
(|{\bf r}-{\bf S}_{\alpha}|G) Y_{lm}^{\ast} (\hat{{\bf G}}) Y_{lm} (\bf {\widehat{r-S_{\alpha}}})\;.
\end{equation}
\subsection{Local orbitals}
The extension of the LAPW basis set by localized orbitals \cite{Singh:91} was
introduced in order to describe semi-core states, i.e. those low-lying states which reach out of the atomic sphere, on the same footing as valence states. The corresponding basis functions reads inside the atomic sphere 
\begin{eqnarray}
& & \phi_{{\bf {k}},lm}^{LO} ({\bf S}_{\alpha} + {\bf r}) = \nonumber \\ 
& &
\left[
        A_{lm}^{\alpha}({\bf {k+G}}_{lm}^{LO})       u_{l}^{\alpha}(r,E_l) +
        B_{lm}^{\alpha}({\bf {k+G}}_{lm}^{LO}) \dot{u}_{l}^{\alpha}(r,E_l) +
        C_{lm}^{\alpha}({\bf {k+G}}_{lm}^{LO})       u_{l}^{\alpha}(r,E_{l}^{LO}) \right] 
Y_{l,m} ({\bf{{\hat r}}}) \nonumber \\
\label{basis_LO}
\end{eqnarray}
and is zero outside. The coefficients $A_{lm}^{\alpha}({\bf {k+G}}_l^{LO})$,
$B_{lm}^{\alpha}({\bf {k+G}}_l^{LO})$, and $C_{lm}^{\alpha}({\bf {k+G}}_l^{LO})$ are
determined by choosing $\phi_{{\bf {k}},lm}^{LO}$ and its spatial derivative to vanish 
at the sphere boundary, and by the normalization condition for 
$\phi_{{\bf {k}},lm}^{LO}$. The additional trial energy $E_{l}^{LO}$ corresponds
to the energy of the semi-core state, and for each localized basis function 
one specific ${\bf G}_{lm}^{LO}$ is chosen.\cite{Singh:91}
The wavefunction can then be written as
\begin{equation}
\Psi_{n {\bf k}}({\bf r}) =
\sum_{{\bf G}}      C_{n{\bf k}}({\bf G})               \phi_{{\bf {k+G}}}({\bf r}) +
\sum_{lm}             C_{n{\bf k},lm}^{LO}({\bf G}_{lm}^{LO}) \phi_{{\bf {k}},lm}^{LO}({\bf r})
\label{ansatz_LO}
\end{equation}
Analogously to Eqs. (\ref{Phi_x+y}--{\ref{Phi_z}) matrix elements between LAPW's and LO's
\begin{eqnarray}
^{\alpha}\Phi_{{{\bf k}lm},\bf {k+G}}^{x+iy} & \equiv &
\langle \phi_{{\bf k},lm}^{LO} ( {\bf S}_{\alpha} + {\bf r} )
\left| \partial x + i\partial y \right| 
\phi_{\bf {k+G}} ( {\bf S}_{\alpha} + {\bf r} ) \rangle 
\label{Phi_LO_x+y}\\
^{\alpha}\Phi_{{{\bf k}lm},\bf {k+G}}^{x-iy} & \equiv &
\langle \phi_{{\bf k},lm}^{LO} ( {\bf S}_{\alpha} + {\bf r} )
\left| \partial x - i\partial y \right| 
\phi_{\bf {k+G}} ( {\bf S}_{\alpha} + {\bf r} ) \rangle 
\label{Phi_LO_x-y}\\
^{\alpha}\Phi_{{{\bf k}lm},\bf {k+G}}^{z} & \equiv &
\langle \phi_{{\bf k},lm}^{LO} ( {\bf S}_{\alpha} + {\bf r} )
\left| \partial z \right| 
\phi_{\bf {k+G}} ( {\bf S}_{\alpha} + {\bf r} ) \rangle
\label{Phi_LO_z}
\end{eqnarray}
and between LO's and LO's have to be defined:
\begin{eqnarray}
^{\alpha}\Phi_{{{\bf k}l^{\prime}m^{\prime}},{\bf k}lm}^{x+iy} & \equiv &
\langle \phi_{{\bf k},l^{\prime}m^{\prime}}^{LO} ( {\bf S}_{\alpha} + {\bf r} )
\left| \partial x + i\partial y \right| 
\phi_{{\bf k},lm}^{LO} ( {\bf S}_{\alpha} + {\bf r} ) \rangle 
\label{Phi_LOLO_x+y}\\
^{\alpha}\Phi_{{{\bf k}l^{\prime}m^{\prime}},{\bf k}lm}^{x-iy} & \equiv &
\langle \phi_{{\bf k},l^{\prime}m^{\prime}}^{LO} ( {\bf S}_{\alpha} + {\bf r} )
\left| \partial x - i\partial y \right| 
\phi_{{\bf k},lm}^{LO} ( {\bf S}_{\alpha} + {\bf r} ) \rangle 
\label{Phi_LOLO_x-y}\\
^{\alpha}\Phi_{{{\bf k}l^{\prime}m^{\prime}},{\bf k}lm}^{z} & \equiv &
\langle \phi_{{\bf k},l^{\prime}m^{\prime}}^{LO} ( {\bf S}_{\alpha} + {\bf r} )
\left| \partial z \right| 
\phi_{{\bf k},lm}^{LO} ( {\bf S}_{\alpha} + {\bf r} ) \rangle 
\label{Phi_LOLO_z}
\end{eqnarray}
The atomic sphere parts of the momentum matrix elements given in Eq. (\ref{ME_alpha}) 
have to be supplemented by contributions from the local orbitals:
\begin{eqnarray}
\langle n^{\prime}{\bf k} \left| \nabla_x \right| n{\bf k} \rangle_{\mathrm{MT}_{\alpha}}
& = &
%
%
\frac{1}{2 \phantom{i}} 
\sum_{G^{\prime},G} 
C_{n^{\prime} {\bf k}}^{\ast} ({\bf G^{\prime}})
\left(
^{\alpha}\Phi_{{\bf {k+G^{\prime}},\bf {k+G}}}^{x+iy}  +
^{\alpha}\Phi_{{\bf {k+G^{\prime}},\bf {k+G}}}^{x-iy} \right)
C_{n {\bf k}} ({\bf G}) \nonumber \\
%
%
& + &
\frac{1}{2 \phantom{i}} 
\sum_{lm,G} 
C^{\ast LO}_{n^{\prime}{\bf k},lm}({\bf G}_{lm}^{LO})
\left(
^{\alpha}\Phi_{{{\bf k}lm},\bf {k+G}}^{x+iy} +
^{\alpha}\Phi_{{{\bf k}lm},\bf {k+G}}^{x-iy} \right)
C_{n {\bf k}} ({\bf G}) \nonumber \\
%
%
& + &
\frac{1}{2 \phantom{i}} 
\sum_{l^{\prime}m^{\prime},lm} 
C^{\ast LO}_{n^{\prime}{\bf k},l^{\prime}m^{\prime}} ({\bf G}_{l^{\prime}m^{\prime}}^{LO})
\left(
^{\alpha}\Phi_{{{\bf k}l^{\prime}m^{\prime}},{\bf k}lm}^{x+iy} +
^{\alpha}\Phi_{{{\bf k}l^{\prime}m^{\prime}},{\bf k}lm}^{x-iy} \right)
C_{n {\bf k},lm}^{LO} ({\bf G}_{lm}^{LO}) \nonumber \\
\label{ME_x_full} \\
\langle n^{\prime}{\bf k} \left| \nabla_y \right| n{\bf k} \rangle_{\mathrm{MT}_{\alpha}}
& = &
%
%
\frac{1}{2i}
\sum_{G^{\prime},G} 
C_{n^{\prime} {\bf k}}^{\ast} ({\bf G^{\prime}})
\left(
^{\alpha}\Phi_{{\bf {k+G^{\prime}},\bf {k+G}}}^{x+iy}  -
^{\alpha}\Phi_{{\bf {k+G^{\prime}},\bf {k+G}}}^{x-iy} \right)
C_{n {\bf k}} ({\bf G}) \nonumber \\
%
%
& + &
\frac{1}{2i}
\sum_{lm,G} 
C^{\ast LO}_{n^{\prime}{\bf k},lm}({\bf G}_{lm}^{LO})
\left(
^{\alpha}\Phi_{{{\bf k}lm},\bf {k+G}}^{x+iy} -
^{\alpha}\Phi_{{{\bf k}lm},\bf {k+G}}^{x-iy} \right)
C_{n {\bf k}} ({\bf G}) \nonumber \\
%
%
& + &
\frac{1}{2i}
\sum_{l^{\prime}m^{\prime},lm} 
C^{\ast LO}_{n^{\prime}{\bf k},l^{\prime}m^{\prime}} ({\bf G}_{l^{\prime}m^{\prime}}^{LO})
\left(
^{\alpha}\Phi_{{{\bf k}l^{\prime}m^{\prime}},{\bf k}lm}^{x+iy} -
^{\alpha}\Phi_{{{\bf k}l^{\prime}m^{\prime}},{\bf k}lm}^{x-iy} \right)
C_{n {\bf k},lm}^{LO} ({\bf G}_{lm}^{LO})
\nonumber\\
\label{ME_y_full} 
\\
\langle n^{\prime}{\bf k} \left| \nabla_y \right| n{\bf k} \rangle_{\mathrm{MT}_{\alpha}}
& = &
%
%
\phantom{\frac{1}{2i}}
\sum_{G^{\prime},G} 
C_{n^{\prime} {\bf k}}^{\ast} ({\bf G^{\prime}}) \phantom{(}
^{\alpha}\Phi_{{\bf {k+G^{\prime}},\bf {k+G}}}^{z}  \phantom{)}
C_{n {\bf k}} ({\bf G}) \nonumber \\
%
%
& + &
\phantom{\frac{1}{2i}}
\sum_{lm,G} 
C^{\ast LO}_{n^{\prime}{\bf k},lm}({\bf G}_{lm}^{LO}) \phantom{(}
^{\alpha}\Phi_{{{\bf k}lm},\bf {k+G}}^{z} \phantom{)}
C_{n {\bf k}} ({\bf G}) \nonumber \\
%
%
& + &
\phantom{\frac{1}{2i}}
\sum_{l^{\prime}m^{\prime},lm} 
C^{\ast LO}_{n^{\prime}{\bf k},l^{\prime}m^{\prime}} ({\bf G}_{l^{\prime}m^{\prime}}^{LO})
\phantom{(}
^{\alpha}\Phi_{{{\bf k}l^{\prime}m^{\prime}},{\bf k}lm}^{z} \phantom{)}
C_{n {\bf k},lm}^{LO} ({\bf G}_{lm}^{LO})
\label{ME_z_full}
\end{eqnarray}
The evaluation of Eqs. (\ref{Phi_LO_x+y}-\ref{Phi_LOLO_z}) is straightforward analogous to Eqs. (\ref{Phi_x+iy_final}-\ref{Phi_z_final}).

\end{document}